\documentclass[twocolumn,showpacs,preprintnumbers, amsmath,amssymb,amsfonts, final, a4paper,aps,
citeautoscript,footinbib,prb,superscriptaddress]{revtex4}
\usepackage{times}
\usepackage{graphicx}
\usepackage[latin1]{inputenc}
\usepackage{mathrsfs}
\usepackage[a]{esvect}
\renewcommand{\vec}{\vv}
  
\graphicspath{{figures/}}

\newcommand{\ZZ}{\mathbb{Z}}

\newcommand{\ds}{\displaystyle}

\newcommand{\cV}{{\cal V}}

\newcommand{\pkm}{\varphi_{k}}

\newcommand{\pkp}{\varphi_{k+1}}

\newcommand{\pd}{\partial}
\newcommand{\tC}{\widetilde{C}}
\newcommand{\tS}{\widetilde{S}}

\begin{document}
\title{Exotic Quantum Criticality in One-dimensional Coupled Dipolar Bosons Tubes}
 \author{P. Lecheminant} 
\affiliation{Laboratoire de Physique Th\'eorique et Mod\'elisation,
  CNRS UMR 8089, Universit\'e de Cergy-Pontoise, Site de Saint-Martin,
  F-95300 Cergy-Pontoise Cedex, France}
\author{H. Nonne} 
\affiliation{Department of Physics, Technion, Haifa 32000, Israel}

\date{\today}
\pacs{{71.10.Pm}, 
{05.30.Jp} 
}

\begin{abstract}
The competition between intertube hopping processes and density-density interactions
is investigated in one-dimensional quantum dipolar bosons systems of $N$ coupled tubes at zero temperature.
Using a phenomenological bosonization approach, we show that the resulting
competition leads to an exotic quantum phase transition described 
by a U(1) $\times$ $\ZZ_N$ conformal field theory with a fractional central charge.
The emerging $\ZZ_N$  parafermionic critical degrees of freedom are highly nontrivial in terms of the original atoms or
polar molecules of the model. We further determine the main physical properties of the quantum critical
point in a double-tube system which has central charge $c=3/2$. In triple-tube systems, 
we show that the competition between the two antagonistic processes is related to the physics of the
two-dimensional $\ZZ_3$ chiral Potts model.
This work opens the possibility to study the exotic properties of the $\ZZ_N$ parafermions
in the context of ultracold quantum Bose gases.
  \end{abstract}

\maketitle

\section{Introduction}
Ultracold atomic systems have become one of the most experimentally flexible
systems to study strongly correlated physics. 
The possibility of tuning the contact inter-particle interactions by varying
the scattering length through Feshbach resonances 
has allowed the observation of many interesting phenomena like
the BEC-BCS crossover in atomic Fermi gas \cite{review}.

On top of this wide tunability of the interaction strength,
the possibility of controlling the shape of interactions  through the long-range
and anisotropic nature of dipole-dipole interactions has recently attracted
considerable attention. In particular, dipolar interactions with long-range anisotropic character
have been observed in chromium atoms by exploiting its large magnetic moments \cite{reviewdipole}. 
An alternative route is to consider heteronuclear polar molecules with large electric dipole moments
associated with their rotational excitations \cite{ye,aikawa}. The resulting long-range interactions between such 
polar molecules can be tuned using dc and ac electric fields \cite{micheli07,gorshkov08}.
Dipole-dipole interactions can then be much stronger than the superexchange interactions between
ultracold atoms.
This opens an avenue to realize a plethora of interesting quantum phases governed by long-range
interactions. On the one hand, a variety of exotic phases have already been predicted to occur
in dipolar quantum gases such as topological p-wave fermionic superfluids \cite{cooper} and quantum nematic fluids
\cite{quintanilla,fradkindip}.  On the other hand, as shown recently in Ref.  \onlinecite{gorshkov}, 
ultracold polar molecules become promising candidates  for the simulation of condensed matter phenomena 
and could provide a robust toolbox for quantum information processing.

In this paper, we investigate the emergence of an exotic quantum criticality in 
coupled one-dimensional (1D) quantum bosons tubes with dipole-dipole interactions.
The standard quantum critical behavior of  1D quantum bosons is 
the well-known Luttinger universality class \cite{haldane,giamarchi,Cazalilla,ReviewBose}. 
Its low-energy properties are governed by a relativistic free massless boson field $\Phi$
with a sound velocity $v$ and a Luttinger parameter $K$ which depend on the 
density and interactions of the underlying 1D quantum Bose gas. A hallmark of the
Luttinger phenomenology is the power-law decay of 
the physical quantities with non-universal exponents related to the Luttinger parameter $K$.
Such universality class can in turn be viewed as a conformal field theory (CFT) 
with U(1) global continuous symmetry and central charge $c=1$ associated to the
gapless free-bosonic mode $\Phi$ \cite{dms,bookboso}.
A more complicate quantum-critical behavior has been found recently with ultracold bosons
with the possible realization of the 2D Ising universality class 
\cite{sachdevIsing,berg,diehl,joe,altman}.
In that case, the criticality emerges at a quantum critical point characterized by a $\ZZ_2$ CFT with central charge $c=1/2$.
The resulting gapless degree of freedom is a free massless Majorana (real) fermion which 
is half  of a Dirac (complex) fermion \cite{dms,bookboso}. 
Here, in the context of coupled 1D dipolar bosons, we analyse the stabilization of exotic quantum critical
points which are described by the SU(2)$_N$ CFT with central charge $c=3N/(N+2)$ \cite{knizhnik}.
Equivalently, the emergent quantum criticality can be captured by the product of a conventional
U(1) CFT with the $\ZZ_N$ parafermions CFT, with central charge $c=2(N-1)/(N+2)$, which
is known to govern the critical properties of the 2D $\ZZ_N$ generalization 
of Ising models \cite{zamolo,gepner}. 
Such nontrivial quantum criticality for  $N>2$ cannot be described by a 
simple free-field theory in terms of 1D gapless bosons or fermions
as it is the case for the Luttinger and Ising universality classes. In this respect, the resulting critical degrees of
freedom are highly nonlocal  with respect to the original atoms or molecules.
Recently, the $\ZZ_N$ parafermions CFT has attracted a lot of interest in the context
of non-Abelian fractional quantum Hall states.  The so-called Read-Rezayi states are described in terms
of the $\ZZ_N$ CFT and their excitations display non-Abelian statistics which stem from 
the parafermionic behavior of these states \cite{read}.
Our work opens the possible realization of this exotic 
CFT with fractional central charge in the context of ultracold atoms or polar molecules.

\begin{figure}
     \begin{center}
     \includegraphics[width=0.99\columnwidth,clip]{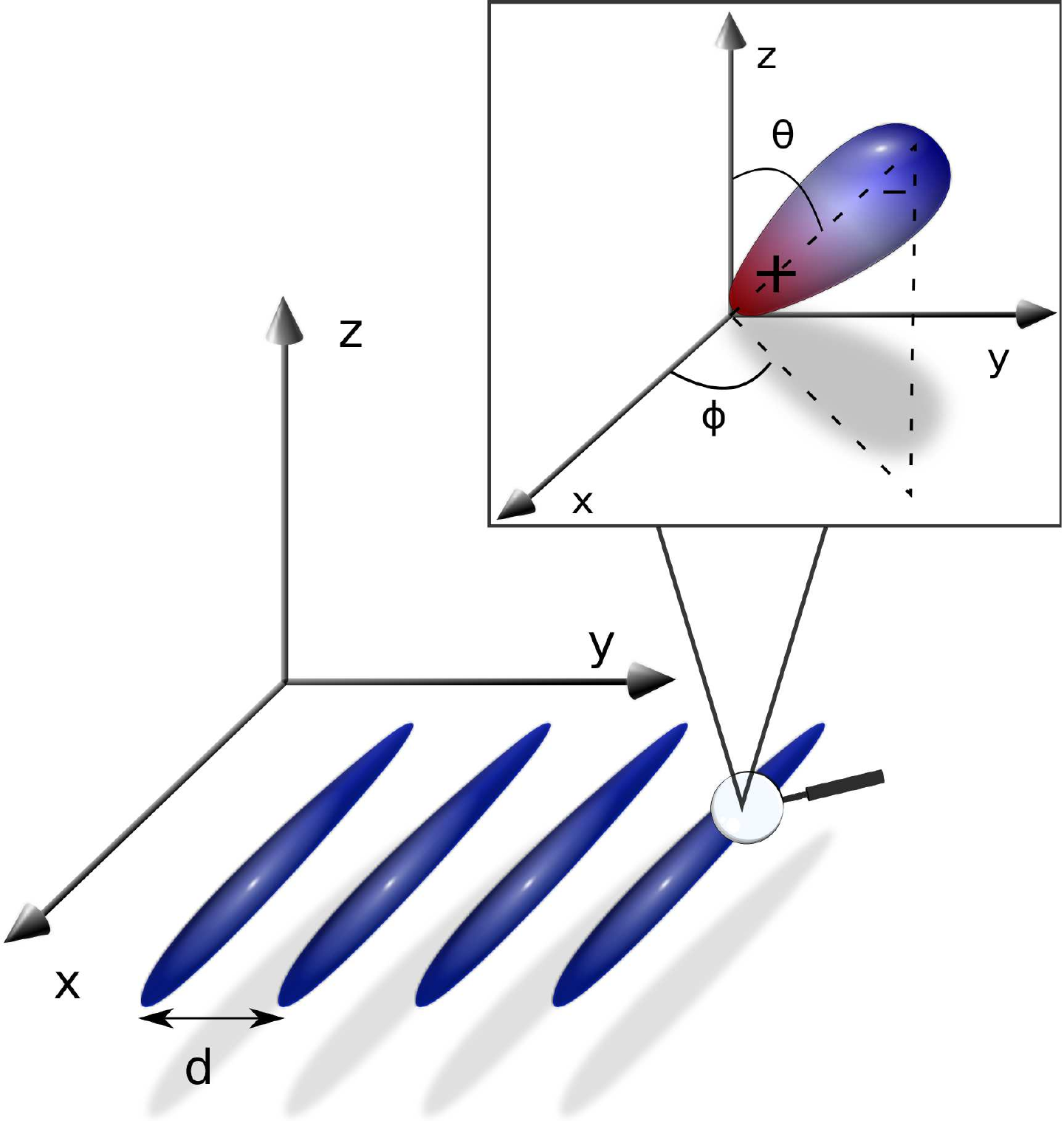}
\end{center}
\caption{(color online) Planar array of 1D tubes of dipolar boson gas positioned along the $x$ axis. The dipoles are all aligned along an additional external field.}
\label{fig:bosonsN}
\end{figure}
The experimental setup that allows to create a system of coupled 1D dipolar bosons tubes, is similar to the one described in Refs. \onlinecite{fellows,arguelles,kollath,wang,dalmonte,zinner}.
A planar array of 1D tubes of dipolar boson gas, as shown in Fig. \ref{fig:bosonsN}, is formed by an anisotropic 2D optical lattice in the $z-y$ plane. The strong confinement in the $z$-direction completely suppresses the hopping in that direction, while we allow a weak hopping between the tubes along the $y$-direction. 
 Neglecting the inhomogeneity of the trapping potential, the system is thus a stack of 2D arrays, containing identical (straight) tubes of bosons, all oriented along the $x$-axis and separated by a distance $d = \lambda/2$ in the $y$-direction
 ($\lambda$ being the wavelength of the laser) . 
 The dipole moments of the bosons are aligned by an additional external field which is electric in the case of polar molecules
or magnetic for chromium atoms. The dipole-dipole interaction strength can 
be tuned by using a rotating field, which induces a precession of the dipoles around the $z$-axis \cite{reviewdipole,pfauPRL}.

The resulting effective Hamiltonian for this system can be written as:
 \begin{equation}
 {\cal H} = {\cal H}_{\rm intra} + {\cal H}_{\rm inter}.
 \label{ham}
\end{equation}
${\cal H}_{\rm intra}$ is the Hamiltonian describing the bosons' motion and interactions within the same tube:
\begin{eqnarray}
&&{\cal H}_{\rm intra} = \sum_{n=1}^{N} \int dx \; \Psi^{\dagger}_n \left(x \right) ( - \frac{\hbar^2}{2 M}  \frac{d^2}{dx^2}) 
\Psi_n\left(x \right)
\nonumber\\
	 &&+  \sum_{n=1}^{N} \int dx_1 dx_2 \; V^{\rm intra}_{dd}\left(x_1 - x_2\right) 
	 \rho_n\left(x_1 \right) \rho_n\left( x_2\right), 
\label{hamintra}
\end{eqnarray}	 
where $\Psi^{\dagger}_n \left(x \right)$ is the bosonic operator that creates a dipole at position $x$ in tube $n$; $\rho_n(x) = \Psi^{\dagger}_n(x) \Psi_n(x)$ is the density operator and normal ordering is implicit throughout this paper.
In Eq. (\ref{hamintra}), $M$ is the mass of the bosons and $V^{\rm intra}_{dd}$
denotes the intratube dipole-dipole interactions. The coupling between the tubes
is composed of the hopping along the $y$ direction, with amplitude $J_\perp$, and the intertube dipole-dipole interactions described by $V^{\rm inter}_{dd}$:
\begin{eqnarray}	 
&&{\cal H}_{\rm inter} = - J_{\perp} \sum_{n=1}^{N-1} \int dx  \left[\Psi^{\dagger}_n \left(x \right)  \Psi_{n+1}\left(x \right)
	+ H.c. \right]
	 \nonumber\\
	&& + \sum_{n=1}^{N-1} \int dx_1 dx_2 \; V^{\rm inter}_{dd}\left(x_1 - x_2\right) 
	 \rho_n\left(x_1 \right) \rho_{n+1}\left( x_2\right).\quad
\label{haminter}
\end{eqnarray}
Here, we only consider coupling between nearest-neighbor tubes as in Ref.~\onlinecite{fellows}. Unless stated otherwise (in
Sec. IV), open-boundary conditions in the transverse direction ($y$) are also assumed in Eq. (\ref{haminter}).
The two terms in Eq. (\ref{haminter}) are of very different nature and can be tuned 
independently: the depth of the optical lattice potential in the $y$ direction controls the inter-tube
hopping amplitude $J_{\perp}$, while $V^{\rm inter}_{dd}$ is 
controlled by the rotating polarizing field. 
 
In this paper, we will investigate the competition between these two sources of coupling, at zero
temperature, and in the 1D regime where the total length in the transverse direction, i.e.
$Nd$, is finite. In this respect, small values of $N$, i.e. $N=2$ or $3$, can be obtained in principle, by 
selective evaporation of the tubes by using electric or magnetic field gradients. Alternatively, 
double-tube system can also be engineered by considering double-well optical lattices \cite{anderlini}. 
We will show that the competition between hopping processes and density-density interactions of Eq. (\ref{haminter})
strongly depends on the value of the Luttinger parameter K of the 1D tubes. For bosons with
contact interactions, K is larger than one and the hopping term turns out to be the leading contribution  \cite{ho}.
A transition occurs, in the Berezinsky-Kosterlitz-Thouless (BKT) universality class, between
the Luttinger phase where the $N$ tubes are decoupled and a superfluid phase with phase-coherence between
neighboring tubes \cite{ho}. In stark contrast to contact interactions, we will show that 
a different scenario happens for 1D dipolar quantum bosons, 
which can have $K <1$ due to the non-local nature of the dipole-dipole interactions \cite{orignac}.
In the regime $K <1$, the hopping and density-density interactions strongly compete and
a quantum phase transition takes place between a superfluid phase and a density or charge-density wave (CDW).
The resulting transition becomes exotic and is described by a U(1) $\times$ $\ZZ_N$ conformal field theory 
with central charge $c=3N/(N+2)$.

The rest of the paper is organized as follows. In Sec. II, we follow a low-energy approach for the model (\ref{ham}) in the general $N$ case, using the phenomenological bosonization of the bosons \cite{haldane,giamarchi,Cazalilla,ReviewBose} 
supplemented by CFT techniques. There, we reveal the existence of a quantum critical point in the 
SU(2)$_N$ universality class, which describes the competition between  hopping and intertube interactions.
The properties of this fixed point and its stability with respect to general perturbations
are further analysed in the special cases $N=2,3$ in Sec. III and IV. In particular, we show that 
the low-energy properties of triple-tube systems are related to the physics of the  $\ZZ_3$ chiral Potts model
in the vicinity of the three-state Potts critical point \cite{ostlundhuse}.
Finally, Sec. V contains our concluding remarks. The paper is supplied with an Appendix which provides some 
important technical details.

\section{Quantum phase transition in the general $N$ tube case}
\label{sec:QPTGeneralCase}
In this section, we will investigate the nature of the quantum phase transition which
stems from the competition between the hopping term  and the intertube density-density
interaction of Eq. (\ref{haminter}) in the general $N$ case.

\subsection{Phenomenological bosonization approach}
Let us first specify the form of the dipole-dipole interactions which appear
in Eqs. (\ref{hamintra}, \ref{haminter}). 
The dipole moments are polarized along a direction ${\bf d} = (\sin \theta \cos \phi, \sin \theta \sin \phi,\cos \theta)$ in the spherical coordinates (see Fig. \ref{fig:bosonsN}).
The energy due to the interaction between two dipoles at relative position $\bf r$ is:
\begin{equation}
V_{\rm dd}\left({\bf r}\right) = \frac{C_{dd}}{4\pi} \frac{r^2 - 3 \left({\bf d} \cdot {\bf r}\right)^2}{r^5},
\label{dipole}
\end{equation}
where the coupling constant is $C_{dd}=p^2/ \epsilon_0$ for electric dipoles of strength $p$, and $C_{dd}=\mu_0 \mu^2$ for magnetic dipoles of strength $\mu$; $\epsilon_0$ and $\mu_0$ are respectively 
the vacuum permittivity and permeability.
For two dipoles in the same tube, the dipole-dipole interaction is then:
\begin{equation}
V^{\rm intra}_{\rm dd}\left(x_1 - x_2\right) = 
\frac{C_{dd}}{4\pi} \frac{ 1 - 3 \sin^2 \theta \cos^2 \phi}{|x_1 - x_2|^3}.
\label{dipoleintra}
\end{equation}
A crucial hypothesis in our work is that the dipole-dipole interaction along the tubes
should be repulsive to reach the regime $K<1$.
A standard choice to get repulsive interactions along and between the tubes 
is to fix $\theta$ to its magic angle value: $\sin \theta = 1/\sqrt{3}$ as in Ref.  \onlinecite{fellows}
for instance. 
However, as we will show thereafter, repulsive intertube dipole-dipole interactions 
are not crucial for the existence of the quantum phase transition we investigate. 
Without loss of generality regarding the nature of the intertube 
interactions, we will consider a simpler choice $\theta = \phi = \pi/2$, where the dipoles are polarized along
the $y$-axis, perpendicular to the tubes axis. We thus get a repulsive intratube interaction:
\begin{equation}
V^{\rm intra}_{\rm dd}\left(x_1 - x_2\right) =
\frac{C_{dd}}{4\pi} \frac{1}{|x_1 - x_2|^3}.
\label{dipoleintra2}
\end{equation}
As in Ref.  \onlinecite{fellows}, we approximate the dipole-dipole interaction between two nearest-neighbor
tubes by its leading contribution:
\begin{equation}
V^{\rm inter}_{\rm dd}\left(x_1 - x_2\right) \simeq
- \frac{C_{dd}}{2\pi d^3} \delta \left( x_1 - x_2\right),
\label{dipoleinter2}
\end{equation}
so that we then have a short-distance intertube attraction.
The later approximation is very convenient to perform a field-theory analysis as in weakly coupled
fermionic or bosonic ladders \cite{giamarchi,bookboso,ReviewBose}.
Hence, we do not treat the full dipole-dipole interaction (\ref{dipole}) in coupled
tubes like in Ref. \onlinecite{bauer}, where the same system was considered, but without intertube hopping.
However, the dipole interaction in quasi-1D (or in 2D) is not fully long-ranged 
as in 3D, in the sense that its volume integral does not diverge with the system size. The crude approximation (\ref{dipoleinter2}) might thus be enough to shed light on
the competition between superfluidity and the CDW ordering in coupled tubes.
In this respect, we will find that model (\ref{ham}) displays a quantum-critical point and we will investigate its stability under generic perturbations -- perturbations that include in particular the longer range terms, neglected in $V^{\rm inter}_{\rm dd}$.
 
With these specifications, we are now in position to study the physical properties
of model (\ref{ham}) by means of the phenomenological bosonization (or harmonic-fluid approach).
It is a powerful method that allows to study the low-energy properties of strongly interacting
1D quantum bosons systems\cite{haldane,giamarchi,Cazalilla,ReviewBose}.
The density operators $ \rho_n\left(x\right), n = 1, \ldots, N$ can be expressed in terms of
 $N$ bosonic quantum fields $\varphi_n(x)$ through: \cite{ReviewBose}
 \begin{equation}
  \rho_n\left(x\right) \simeq 
  \left[ \rho_0 + \frac{1}{\sqrt{\pi}} \partial_x \varphi_n \left(x\right)
\right] \sum_{m=-\infty}^{\infty} \alpha_m
e^{2 i m (\sqrt{\pi} \varphi_n \left(x\right) + \pi \rho_0 x) },
 \label{densboso}
\end{equation}
where $\alpha_m$ are non-universal constants and 
$\rho_0$ is the average density in each tube around which the fluctuations take place.
We introduce the dual bosons $\vartheta_n$ associated to the fields
$\varphi_n $  such that $[\varphi_n(x_1), \partial_x \vartheta_m(x_2)] = 
i \delta_{nm}   \delta (x_1 - x_2)$. They allow us to describe the phase fluctuations of the  boson creation operator 
$\Psi^{\dagger}_n$:
\begin{eqnarray}
\Psi^{\dagger}_n  \left(x \right)  &\simeq &
\left[ \rho_0 + \frac{1}{\sqrt \pi} \partial_x \varphi_n \left(x\right)
\right]^{1/2}    \exp \left( i \sqrt{\pi}\vartheta_n \left(x\right) \right) \nonumber \\
&&
\times\sum_{m=-\infty}^{\infty} \beta_m
e^{2 i m (\sqrt{\pi} \varphi_n \left(x\right) + \pi \rho_0 x) } ,
\label{bosoop}
\end{eqnarray}
$\beta_m$ being non-universal constants. 
In practice, one often only needs the leading parts of the expansions (\ref{densboso},\ref{bosoop}):
\begin{eqnarray}
\rho_n \left(x\right) &\sim&  \rho_0 + \frac{1}{\sqrt \pi} \partial_x \varphi_n \left(x\right)\nonumber\\
&&\qquad+ 2 \rho_0 \cos \left( \sqrt{4\pi} \; \varphi_n \left(x\right) + 2 \pi \rho_0 x \right)
\nonumber \\
\Psi_{n}^{\dagger} \left(x\right) &\sim& 
\sqrt{\rho_0} \; \exp \left( i \sqrt{\pi} \vartheta_n \left(x\right) \right) .
\label{bosodico}
\end{eqnarray}
After investigation of the effects of these leading parts, the higher harmonics and less
relevant terms can then be studied a posteriori.

In the absence of intertube coupling, the low-energy properties of the Hamiltonian (\ref{hamintra}) are described
by the Tomonaga-Luttinger model: \cite{orignac}
\begin{equation}
{\cal H}_{\rm intra} = \sum_{n=1}^{N}  \int dx \; \frac{v}{2} \left[ K \left(\partial_x \varphi_n 
\right)^2 +  \frac{1}{K} \left(\partial_x \vartheta_n \right)^2 \right],
\label{hamluttinger}
\end{equation}	 
where the velocity $v$, and the Luttinger parameter $K$ depend on the microscopic details
of the original model. The Luttinger parameter has been determined numerically as function
of the dipole coupling constant $C_{dd}$  and the average density $\rho_0$ \cite{dalmonte,orignac}. In sharp contrast 
to bosonic gases with contact interactions where $K \ge 1$, $K$ can be much smaller than 1 
for dipolar interactions. 

The leading-part of the intertube interaction (\ref{haminter}) can be expressed in terms of the boson fields
$\varphi_n $ and $\vartheta_n$ using Eq. (\ref{bosodico}):
\begin{eqnarray}
{\cal H}_{\rm inter} &=& \int dx \; \sum_{n=1}^{N-1} \left[ - \frac{g_{\perp}}{a^2} 
\cos \left( \sqrt{\pi} \left( \vartheta_{n+1} -  \vartheta_{n} \right) \right) \right.
\nonumber \\
&-& \left.  \frac{g_{d}}{a^2} 
\cos \left( \sqrt{4\pi} \left( \varphi_{n+1} -  \varphi_{n} \right) \right) \right]
\nonumber \\
&+& \lambda  \int dx \; \sum_{n=1}^{N-1} \partial_x \varphi_{n+1} \partial_x \varphi_{n} ,
\label{bosointer}
\end{eqnarray}
$a \sim 1/\rho_0$ being the short distance cut off (the average interparticle distance). For the attractive dipole-dipole interaction (\ref{dipoleinter2}), the coupling constants
read as follows :
\begin{eqnarray}
g_{\perp} &=& 2 \rho_0 a^2 J_{\perp} \nonumber \\
g_d &=& \frac{C_{dd}  \rho^2_0 a^2 }{\pi d^3} 
\nonumber \\
  \lambda &=&   - \frac{C_{dd}}{2 \pi^2 d^3} .
\label{couplings}
\end{eqnarray}
At this point, it is worth noting that the sign of the two cosine operators of Eq. (\ref{bosointer})
can be changed by the canonical transformation on the bosonic fields:
 \begin{eqnarray}
 \varphi_{2n}  &\rightarrow& \varphi_{2n} + \frac{\sqrt{\pi}}{4}, \;  \; \varphi_{2n+1}  \rightarrow \varphi_{2n+1}
 \nonumber \\
  \vartheta_{2n}  &\rightarrow& \vartheta_{2n} + \sqrt{\pi},  \;  \; \vartheta_{2n+1}  \rightarrow \vartheta_{2n+1} ,
 \label{cantrans}
\end{eqnarray}
so that the case of repulsive intertube interactions can be treated
on the same footing by using the transformation (\ref{cantrans}).

Let us discuss the different terms that appear in the interacting Hamiltonian (\ref{bosointer}). They are of very different nature.
The forward scattering term with the coupling constant $\lambda$ will  renormalize the velocity and Luttinger parameter of the bosonic modes. 
Now, the structure of the remaining terms is intriguing.
They describe the competition between the bosonic fields $\varphi_{n}$ and their dual fields 
$\vartheta_n$. The low-energy effective Hamiltonian (\ref{bosointer}) 
is, in fact, similar to the competition between superconductivity and CDW
in spin-gap electronic ladder systems \cite{tsvelik,fradkin}, the two first terms in Eq. (\ref{bosointer}) 
being respectively the analogues of the Josephson and CDW interactions.
Their scaling dimensions with respect to the Luttinger-liquid fixed point (\ref{hamluttinger}) are
respectively $\Delta_{\perp} = 1/(2K) $, $\Delta_{d} = 2K $.
The hopping term is thus a relevant perturbation when $K > 1/4$, and the density-density
 term, when $K<1$.
For contact interactions, $K>1$, so the density-density term is irrelevant 
and a BKT transition, driven by the hopping term, leads to a superfluid phase with phase coherence 
between the tubes\cite{ho}.
For $1/4 < K<1$, which is the regime we are focusing on here,  the two 
perturbations are strongly relevant competing perturbations. A different 
quantum phase transition emerges from this competition. In this case,
the infrared (IR) properties of the system is governed by the effective Hamiltonian: 
\begin{eqnarray}
{\cal H}_{\rm eff} &=& 
 \sum_{n=1}^{N}  \int dx \; \frac{v}{2} \left[ K \left(\partial_x \varphi_n 
\right)^2 +  \frac{1}{K} \left(\partial_x \vartheta_n \right)^2 \right]
\nonumber \\
&-& \int dx \; \sum_{n=1}^{N-1} \left[ \frac{g_{\perp}}{a^2} 
\cos \left( \sqrt{\pi} \left( \vartheta_{n+1} -  \vartheta_{n} \right) \right) \right.
\nonumber \\
&+& \left.  \frac{g_{d}}{a^2} 
\cos \left( \sqrt{4\pi} \left( \varphi_{n+1} -  \varphi_{n} \right) \right) \right],
\label{hameff}
\end{eqnarray}
where the marginal forward scattering term has been discarded.

Our strategy thus consists in, first investigating the physical properties of model
(\ref{hameff}),
and then, analysing the main effect of less relevant perturbations omitted in (\ref{hameff}), such as the forward scattering
and higher harmonics of the bosonized representations (\ref{densboso},\ref{bosoop}).

One can anticipate on symmetry grounds that the strongly relevant terms of model  (\ref{hameff}) will 
not introduce a spectral gap for all degrees of freedom. Indeed, it is straightforward
to see that the model is invariant under two independent global U(1) transformations:
\begin{eqnarray}
 \varphi_{n} &\rightarrow&  \varphi_{n} + \alpha \nonumber \\
 \vartheta_{n} &\rightarrow&  \vartheta_{n} + \beta ,
 \label{U1sym}
\end{eqnarray}
$\alpha, \beta$ being real numbers. Introducing left and right components of the bosonic fields
$\varphi_{nL,R} = (\varphi_{n} \pm \vartheta_{n})/2$, the transformation (\ref{U1sym}) 
gives a U(1)$_L$ $\times$ U(1)$_R$ global continuous symmetry of model (\ref{hameff}).
We thus expect the model to be gapless with one bosonic field protected by the symmetry  (\ref{U1sym}).

One way to reveal the emergence of this bosonic mode
is to single it out by performing a change of basis on the bosons fields. In this respect, we thus introduce the bosonic field
$\Phi_{0R,L}$, and $N-1$ other fields $\Phi_{l R,L}$
($l = 1, \ldots, N-1$) as follows:
\begin{eqnarray}
\Phi_{0R(L)} &=& \frac{1}{\sqrt{N}}\left(
\varphi_1 + \ldots + \varphi_N \right)_{R(L)}
\nonumber \\
\Phi_{l R(L)} &=& \frac{1}{\sqrt{l(l+1)}}\left(
\varphi_1 + \ldots + \varphi_l - l  \varphi_{l+1}\right)_{R(L)} .
\label{SUNbasis}
\end{eqnarray}
Note that in the context of condensed matter physics, the boson $\Phi_{0R,L}$ would be called the charge boson since it is simply the sum of all bosons and describes  fluctuations of the total charge. 
 The inverse transformation can be easily found; we do not however need its explicit
form but only some of its general properties:
\begin{equation}
\varphi_{nR(L)} = \frac{1}{\sqrt{N}} \Phi_{0R(L)} + {\bf v}^{(n)} \cdot {\bf \Phi}_{R(L)},
\label{invtrans}
\end{equation}
where ${\bf \Phi}$ is the $N-1$ dimensional vector with components $\Phi_{l}$ ($l = 1, \ldots, N-1$)
and the $N$ vectors ${\bf v}^{(n)}$ ($n = 1, \ldots, N$) satisfy the following relations:
\begin{eqnarray}
{\bf v}^{(n)} \cdot {\bf v}^{(m)} &=& \delta_{nm} - \frac{1}{N}
\nonumber \\
\sum_{n=1}^{N} {\bf v}^{(n)} &=& {\bf 0}
\nonumber \\
\sum_{n=1}^{N} v^{(n)}_p v^{(n)}_q &=& \delta_{pq} ,
\label{SUNrelations}
\end{eqnarray}
with $p,q = 1, \ldots, N-1$.
In the new basis, the effective Hamiltonian (\ref{hameff}) simplifies as follows:
\begin{eqnarray}
{\cal H}_{\rm eff} &=& 
  \int dx \; \frac{v}{2} \left[ K \left(\partial_x \Phi_0 
\right)^2 +  \frac{1}{K} \left(\partial_x \Theta_0 \right)^2 \right]
\nonumber \\
&+& 
\int dx \; \frac{v}{2} \left[ K \left(\partial_x {\bf \Phi}
\right)^2 +  \frac{1}{K} \left(\partial_x {\bf \Theta} \right)^2 \right]
\nonumber \\
&-& \int dx \; \sum_{n=1}^{N-1} \left[ \frac{g_{\perp}}{a^2} 
\cos \left( \sqrt{\pi} \left( {\bf v}^{(n+1)} - {\bf v}^{(n)} \right) \cdot {\bf \Theta} \right) \right.
\nonumber \\
&+& \left.  \frac{g_{d}}{a^2} 
\cos \left( \sqrt{4\pi} \left( {\bf v}^{(n+1)} - {\bf v}^{(n)} \right) \cdot {\bf \Phi} \right)
 \right] ,
\label{hameffbis}
\end{eqnarray}
where $\Theta_0$ and ${\bf \Theta}$ are respectively the dual fields
of $\Phi_0$ and ${\bf \Phi}$.
It is interesting to observe that the low-energy effective theory (\ref{hameffbis}) has been 
found in total different contexts. On the one hand, the deconfining phase transition of 
the 2+1-dimensional SU($N$) Georgi-Glashow model is controlled by model  (\ref{hameffbis}) \cite{phle}.
On the other hand, as shown recently, it describes the Read-Rezayi sequence
of non-Abelian fractional quantum Hall states  \cite{kane}.
We explicitly observe, from Eq. (\ref{hameffbis}),  that the $\Phi_0$ field decouples from the interaction terms and remains
gapless. We now need to analyse the fate of the remaining degrees of freedom. The low-energy properties of the model depend on the value of $K$
which controls the competition between the ${\bf \Phi}$ and ${\bf \Theta}$ fields 
of Eq. (\ref{hameffbis}).

When $1/2<K<1$, the most relevant contribution is the dual field ${\bf \Theta}$  term
with scaling dimension $\Delta_{\perp} = 1/(2K) $.
This field is pinned onto the minima of its sine-Gordon model in Eq. (\ref{hameffbis})
with coupling constant $g_{\perp} > 0$: $\langle {\bf \Theta} \rangle = {\bf 0}$. 
In contrast, the bosonic ${\bf \Phi}$ field is a strongly
fluctuating field being dual to ${\bf \Theta}$. The resulting phase is a superfluid phase
which is described by the following leading asymptotics of equal-time correlation functions:
\begin{eqnarray}
\langle  \Psi^{\dagger}_n \left(x \right) \Psi_m \left(0 \right) \rangle & \sim& x^{-1/(2NK)} 
\nonumber \\
\langle  \rho_n \left(x \right) \rho_m \left(0 \right) \rangle &  \sim& \rho^2_0 
- \frac{K}{2 N \pi^2 x^2} .
\label{superfluid}
\end{eqnarray}
In particular, the $2 \pi \rho_0$ oscillations of the density $\rho_n$ are short-ranged.

When $K<1/2$, the most relevant contribution of Eq. (\ref{hameffbis})
is now the term with the bosonic field ${\bf \Phi}$, with scaling dimension $\Delta_{d} = 2K$. The bosonic field ${\bf \Phi}$ is pinned onto the minima of its sine-Gordon model with coupling constant $g_{d} > 0$:
$\langle {\bf \Phi} \rangle = {\bf 0}$. 
In this phase, the bosonic ${\bf \Theta}$ field is a strongly
fluctuating field. The resulting phase is a CDW phase, or a density-wave phase, where the intratube density are all 
in phase:
\begin{eqnarray}
\langle  \Psi^{\dagger}_n \left(x \right) \Psi_m \left(0 \right) \rangle & \sim& e^{- x/\xi}
\nonumber \\
\langle  \rho_n \left(x \right) \rho_m \left(0 \right) \rangle &  \sim& \rho^2_0  + A \frac{\cos
\left(2 \pi \rho_0 x \right)}{x^{2K/N}} ,
\label{cdw}
\end{eqnarray}
$\xi$ being the correlation length which stems from the spectral
gap opened by the relevant perturbation of Eq. (\ref{hameffbis}) 
with coupling constant $g_{d}$, and $A$ is a non-universal constant.
While the bosonic creation operator has short-ranged correlations,
this phase allows the formation between the tubes of a molecular superfluid instability made of $N$ bosons.
Indeed, using the bosonization approach (\ref{bosodico}), we find that the operator 
$M^{\dagger} = \Psi^{\dagger}_1 \Psi^{\dagger}_2 \ldots 
\Psi^{\dagger}_N$ is described by
$M^{\dagger}  \sim  e^{ i \sqrt{\pi} \sum_{n=1}^{N}\vartheta_n } = e^{ i \sqrt{\pi N}\Theta_0 }$.
We deduce thus the leading asymptotics of molecular superfluid correlation function:
\begin{equation}
\langle  M^{\dagger} \left(x \right) M \left(0 \right) \rangle  \sim x^{-N/(2K)}.
\label{molecularsuperfluid}
\end{equation}
This instability dominates the density wave (\ref{cdw}) only if $K > N/2$. 
The situation is in close parallel to the formation of molecular fermionic superfluid instability 
in 1D multicomponent cold fermionic atoms \cite{phlefermions,wu,roux,azaria,hofstetter,dukelsky}.

For decoupled tubes, it is well known that the superfluid correlations are dominant
when $K>1/2$ while the CDW is the leading instability for $K<1/2$ \cite{giamarchi}.
There is a smooth cross-over at $K=1/2$ between these two regions.
But as we can see from Eqs. (\ref{superfluid}, \ref{cdw}), here the situation is in stark contrast to the decoupled tubes case, since 
$K=1/2$ marks the onset of a zero-temperature quantum phase transition.
At $K=1/2$, the competition between the bosonic fields
and their dual is maximal since the two strongly relevant
perturbations have the same scaling dimension $\Delta_\perp=\Delta_d=1$:
\begin{eqnarray}
{\cal H}_{\rm eff} &=& 
 \sum_{n=1}^{N}  \int dx \; \frac{v}{2} \left[ \left(\partial_x \varphi_n 
\right)^2 +  \left(\partial_x \vartheta_n \right)^2 \right]
\nonumber \\
&-& \int dx \; \sum_{n=1}^{N-1} \left[ \frac{g_{\perp}}{a^2} 
\cos \left( \sqrt{2\pi} \left( \vartheta_{n+1} -  \vartheta_{n} \right) \right) \right.
\nonumber \\
&+& \left.  \frac{g_{d}}{a^2} 
\cos \left( \sqrt{2\pi} \left( \varphi_{n+1} -  \varphi_{n} \right) \right) \right] .
\label{hamefftrans}
\end{eqnarray}

\subsection{Non-Abelian symmetry}

The effective Hamiltonian (\ref{hamefftrans}) at $K=1/2$ displays a hidden SU(2)
symmetry when $g_{\perp} = g_{d}$ \cite{tsvelik,fradkin}. In this respect, it is worth introducing 
the following family of SU(2) matrices:
\begin{eqnarray}
g_n  = \frac{1}{\sqrt{2}}
\left(
\begin{array}{lccr}
 e^{-i \sqrt{2\pi} {\varphi}_n}  &
i  e^{-i \sqrt{2\pi} {\vartheta}_n}  \\
i  e^{i \sqrt{2\pi} {\vartheta}_n}   &
 e^{i \sqrt{2\pi} {\varphi}_n} 
\end{array} \right) ,
\label{gbos}
\end{eqnarray}
with $n=1, \ldots, N$. 
The interaction part, $V_{\rm eff}$, of model (\ref{hamefftrans}) can be expressed in terms of these matrices:
\begin{eqnarray}
V_{\rm eff} &=&  \int dx \;  \frac{1}{2 a^2} \left[ - \left(g_{\perp} + g_{d} \right) 
{\rm Tr} \left(g^{\dagger}_{n+1} g_{n} \right) \right.
\nonumber \\
&+& \left. \left(g_{\perp} - g_{d} \right) 
{\rm Tr} \left(g^{\dagger}_{n+1} \sigma_z g_{n} \sigma_z \right)\right]  ,
\label{SU2potent}
\end{eqnarray}
$\sigma_z$ being the third Pauli matrix.
We conclude that the interaction term (\ref{SU2potent}) displays a non-Abelian symmetry
SU(2)$_L$ $\times$ SU(2)$_R$ for $g_{\perp} = g_{d}$ since
it is then invariant under the symmetry:
$g_{n} \rightarrow U g_{n} V$, $U$ and $V$ being two {\sl independent} SU(2) matrices.
When $g_{\perp} \ne g_{d}$, due to the last term of Eq. (\ref{SU2potent}), 
this non-Abelian symmetry is broken down to 
U(1)$_L$ $\times$ U(1)$_R$  in full agreement with  Eq. (\ref{U1sym}).

It is then natural to expect that the quantum phase transition between 
the superfluid and CDW phase occurs precisely at $g_{\perp} = g_{d}$ 
where the effective Hamiltonian (\ref{hamefftrans}) enjoys an enlarged
SU(2)$_L$ $\times$ SU(2)$_R$ global symmetry as well as a self-dual symmetry
${\varphi}_n \leftrightarrow  {\vartheta}_n$:
\begin{eqnarray}
{\cal H}^{*} &=& 
 \sum_{n=1}^{N}  \int dx \; \frac{v}{2} \left[ \left(\partial_x \varphi_n 
\right)^2 +  \left(\partial_x \vartheta_n \right)^2 \right]
\nonumber \\
&-& \frac{g_{d}}{a^2}  \int dx \; \sum_{n=1}^{N-1} \left[ 
\cos \left( \sqrt{2\pi} \left( \vartheta_{n+1} -  \vartheta_{n} \right) \right) \right.
\nonumber \\
&+& \left.  
\cos \left( \sqrt{2\pi} \left( \varphi_{n+1} -  \varphi_{n} \right) \right) \right] .
\label{sdsg}
\end{eqnarray}
The existence of this global SU(2)$_L$ $\times$ SU(2)$_R$  continuous symmetry 
leads us to expect that model (\ref{sdsg}) displays quantum critical properties which
are governed by some CFT. Due to the SU(2)$_L$ $\times$ SU(2)$_R$ invariance, 
the natural candidate is the SU(2)$_k$ CFT with central
charge $c=3k/(k+2)$, where $k$ is some integer to be determined.
In the next subsection, we are going to show non-perturbatively that
the model (\ref{sdsg}) is indeed conformally invariant for all sign of $g_{d}$, and belongs to the SU(2)$_N$
universality class.

\subsection{Conformal embedding approach}
\label{sec:confemb}

Let us now consider the CFTs underlying the physics of our system at the Luttinger-liquid fixed point, i.e. when the intertube interactions are turned off, $g_\perp=g_d=0$. 

At $K=1/2$, each boson field $\varphi_n$ ($n=1, \ldots, N$) is a  compactified boson
with radius $R = 1/ \sqrt{2 \pi}$: $\varphi_n \sim \varphi_n + 2 \pi R$, 
$\vartheta_n \sim \vartheta_n + 2 \pi R$ . 
This equivalence can be viewed as gauge redundancy in the description since the density and 
creation operators (\ref{densboso},\ref{bosoop}) for $K=1/2$ are invariant under that transformation. 
At this special radius, each boson field $\varphi_n$ 
describes an SU(2)$_1$ CFT with central charge $c=1$.
These CFTs are each generated by left and right SU(2)$_1$ currents ${\bf j}_{n R,L}$ which read as follows
in terms of the left-right moving bosons (see Appendix Sec. 1 for more details):
\begin{eqnarray}	
j^{\dagger}_{n R,L} &=& \frac{1}{2 \pi a} \exp( \mp i \sqrt{8\pi}\; \varphi_{n R,L})
\nonumber\\
j^z_{n R,L} &=&  \frac{1}{\sqrt{2 \pi}}   \partial_x \varphi_{n R,L} .
\label{su21current}
\end{eqnarray}	
The SU(2) matrix (\ref{gbos})  is the spin-1/2 SU(2)$_1$ primary operator with scaling dimension $1/2$.
The conformal symmetry of the non-interacting part of model (\ref{sdsg}) is then
$\text{SU(2)}_1 \times \text{SU(2)}_1 \times \ldots
\times \text{SU(2)}_1$ with central charge $N$
(i.e., $N$ gapless bosonic modes). 
The $N$ SU(2)$_1$  currents (\ref{su21current}) can then be combined
to form an SU(2)$_N$  current:
 \begin{eqnarray}	
{\bf I}_{R,L} &=& \sum_{n=1}^{N} {\bf j}_{n R,L} .
\label{su2Ncurrentdef}
\end{eqnarray}	

The next step of the approach
is to go from these $N$ SU(2)$_1$ CFTs to the SU(2)$_N$ CFT with
central charge $3N/(N+2)$. In this respect,
one can use the following conformal embedding to investigate the critical properties
of model (\ref{sdsg}):
\begin{equation}
\text{SU(2)}_1 \times \text{SU(2)}_1 \times \cdots
\times \text{SU(2)}_1 \rightarrow \text{SU(2)}_N \times {\cal G}_N,
\label{embedgenn}
\end{equation}
where  ${\cal G}_N$ is a discrete CFT with
central charge $c_{{\cal G}_N}=N - 3N/(N+2) = N(N - 1)/(N+2)$.
It is important to note that the latter central charge coincides with the sum
of the central charges of the $N-1$ first minimal
models: \cite{phle}
\begin{equation}
c_{{\cal G}_N} = \frac{N \left(N - 1 \right)}{\left(N+2\right)} =
\sum_{m=2}^{N+1} \left( 1 - \frac{6}{m\left(m+1\right)}
\right).
\label{idencharge}
\end{equation} 
The ${\cal G}_N$ CFT is thus related to the product ${\cal M}_3 \times {\cal M}_4 \times \ldots
\times {\cal M}_{N+1}$ where ${\cal M}_p$ denotes the minimal model
series with central charge $c_p = 1 - 6/p(p+1)$; for $p=3,4,5$ they respectively correspond to the Ising, TIM (Tricritical Ising Model), and $\ZZ_3$ Potts CFTs  
\cite{dms}.
The precise identification requires a projection P:
${\cal G}_N \sim P({\cal M}_3 \times {\cal M}_4 \ldots
\times {\cal M}_{N+1})$, which has been described in Ref.
\onlinecite{bul}.

The central point of our analysis is to show that the self-dual perturbation of model (\ref{sdsg}):
\begin{equation}
{\cal V}_{\rm sd} = \sum_{n=1}^{N-1} \cos(\sqrt{2 \pi}(\varphi_{n+1} -  \varphi_{n}))
+  \cos(\sqrt{2 \pi} (\vartheta_{n+1} -  \vartheta_{n})) ,
\label{sdgint}
\end{equation}
does not depend on the SU(2)$_N$ CFT. More precisely,
we show,  in Sec. 2 of the Appendix, that the operator ${\cal V}_{\rm sd}$  
is indeed a singlet under the SU(2)$_N$ CFT and 
a primary operator of the ${\cal G}_N$ CFT with a scaling dimension equal to one.
Hence, since it is a strongly relevant perturbation, ${\cal V}_{\rm sd}$ introduces a spectral gap 
for the discrete ${\cal G}_N$ degrees of freedom, but leaves the SU(2)$_N$ ones intact. 
We thus conclude that model (\ref{sdsg})  
displays critical properties in the SU(2)$_N$ universality class for all signs of $g_{d}$.
 
 However, we are not guaranteed that the critical properties of the initial model
  (\ref{ham}) are controlled by this SU(2)$_N$ fixed point.  One has to investigate
  the stability of the latter fixed point with respect to the marginal forward scattering
  operator of Eq. (\ref{bosointer}) and the  contributions of the decoupled tubes limit such as 
  the higher harmonics of the bosonized description (\ref{densboso},\ref{bosoop}) that we have
  neglected.
 The marginal forward-scattering process breaks the non-abelian SU(2)$_L$ $\times$ SU(2)$_R$ 
 symmetry for $g_{\perp} = g_{d}$ down to U(1)$_L$ $\times$ U(1)$_R$. In this respect,
 the symmetry of the IR fixed point is then U(1) $\times$  $\ZZ_N$ since SU(2)$_N$/U(1) $\sim$
 $\ZZ_N$ \cite{zamolo,gepner}. The latter CFT with central charge 
 $c_N = 2 (N - 1)/(N+2)$  is called the $\ZZ_N$  parafermionic CFT and describes the multicritical properties
 of 2D $\ZZ_N$ generalization of Ising models \cite{zamolo}. The self-duality symmetry,
${\varphi}_n \leftrightarrow  {\vartheta}_n$, of model (\ref{sdsg}) corresponds to 
the Kramers-Wannier (KW) self-duality symmetry of $\ZZ_N$ models. In fact,
the $\ZZ_N$ symmetry can be defined directly in terms of the bosons at $K=1/2$:
\begin{eqnarray}
\vartheta_{n} &\rightarrow&  \vartheta_{n} + \frac{\sqrt{2 \pi}}{N} m
\nonumber \\
 \varphi_{n} &\rightarrow&  \varphi_{n} ,
 \label{Zsymbose}
\end{eqnarray}
with $m=0,1, \dots, N -1$. Using the representation (\ref{bosoop}), this $\ZZ_N$  symmetry has a simple
interpretation in terms of the original bosonic creation operators:
\begin{equation}
\Psi^{\dagger}_n  \left(x \right)  \rightarrow \Psi^{\dagger}_n  \left(x \right)  e^{i \frac{2 \pi}{N} m} .
 \label{Zsym}
\end{equation}

In general, the  SU(2)$_N$  and $\ZZ_N$  fixed points are fragile since they can be destabilized
by several relevant primary operators. Here, the presence of the $\ZZ_N$ symmetry (\ref{Zsymbose})
does protect the system from several relevant perturbations. However, primary operators such as the
thermal operators $\epsilon_k$, with scaling dimensions $\Delta_k = 2 k (k+1)/(N+2)$, 
with $k=1,\ldots [N/2]$ ($[N/2]$ is the integer part of $N/2$),
are invariant under the  $\ZZ_N$ symmetry and might destabilize the IR fixed point.
Fortunately, some of these operators, such as $\epsilon_1$, can be 
killed by a special fine-tuning of the coupling constants by imposing the self-duality symmetry.
Indeed, under the KW symmetry, one has $\epsilon_k \rightarrow - \epsilon_k$, 
and $\epsilon_1$  disappears along the self-dual manifold the model.
As a result, for small values of $N$, the
SU(2)$_N$ or U(1) $\times$  $\ZZ_N$ fixed point is likely to be stable thanks
to the control of the dipole strength and the intertube hopping. 
In contrast, for $N \ge 5$, a strongly relevant self-dual
perturbation $\epsilon_2$ with scaling dimension $12/(N+2)$
will be generated in the ${\mathbb{Z}}_N$ sector.
The resulting field theory which captures the quantum phase transition 
in this case becomes: 
\begin{equation}
{\cal S}_{\rm eff} = {\cal S}_{{\mathbb{Z}}_N}  + \lambda 
\int d^2 x \; \epsilon_2 \left(x\right) , 
\label{efftrans}
\end{equation}
where ${\cal S}_{{\mathbb{Z}}_N}$ stands for the action
of the ${\mathbb{Z}}_N$ CFT.
Model (\ref{efftrans}) turns out to be an integrable deformation
of the ${\mathbb{Z}}_N$ CFT \cite{fateev,dorey}.
The nature of the phase transition depends on 
the sign of the coupling
constant $\lambda$ \cite{fateev,dorey}.
For $\lambda <0$, the field theories (\ref{efftrans}) are
massive and the phase transition is of first-order type.
For $\lambda  > 0$ 
it is known that model (\ref{efftrans})
has a massless renormalization-group flow
onto a BKT  U(1) gapless phase 
with central charge $c=1$. 
In the latter case, there is an intermediate gapless phase between the superfluid and CDW phases for $N \ge 5$.
Adding the contribution of the gapless mode $\Phi_0$, this phase displays extended quantum
criticality with central charge $c=2$.
Unfortunately, our approach does not enable us to fix the sign of the perturbation
of model (\ref{efftrans}) and we cannot discriminate between the two possibilities for the nature of the transition for  $N \ge 5$.
In the following, we will investigate the nature of the quantum phase transition for small values of $N$, i.e.,
$N=2,3$, which respectively correspond to double-tube and triple-tube systems.
 
\section{Double-tube system}
 
 In this section, we determine the zero-temperature physical properties of the double-tube system
 which corresponds to model (\ref{ham}) with $N=2$ (see Fig. (2)).

The harmonic-fluid representation (\ref{hamluttinger}, \ref{bosointer}) simplifies as follows in 
the $N=2$ case:
\begin{eqnarray}
{\cal H}_{N=2} &=&  \int dx \; \frac{v}{2} \left[ K \left(\partial_x \varphi_1 
\right)^2 +  \frac{1}{K} \left(\partial_x \vartheta_1 \right)^2 + 
1 \rightarrow 2\right] \nonumber \\
&-& \int dx \; \left[  \frac{g_{\perp}}{a^2} 
\cos \left( \sqrt{\pi} \left( \vartheta_{2} -  \vartheta_{1} \right) \right) \right.
\nonumber \\
&+& \left.  \frac{g_{d}}{a^2} 
\cos \left( \sqrt{4\pi} \left( \varphi_{2} -  \varphi_{1} \right) \right) \right]
\nonumber \\
&+& \lambda  \int dx \;  \partial_x \varphi_{2} \partial_x \varphi_{1} .
\label{bosoN2}
\end{eqnarray}
The next step is to use the change of basis (\ref{SUNbasis}):
\begin{eqnarray}
\Phi_{0R(L)} &=& \frac{1}{\sqrt{2}}\left(
\varphi_1 + \varphi_2 \right)_{R(L)}
\nonumber \\
\Phi_{1 R(L)} &=& \frac{1}{\sqrt{2}}\left(
\varphi_1  -  \varphi_{2}\right)_{R(L)} ,
\label{SU2basis}
\end{eqnarray}
so that we obtain
\begin{eqnarray}
&& {\cal H}_{N=2} =  \int dx \; \frac{v_0}{2} \left[ K_0 \left(\partial_x \Phi_0 
\right)^2 +  \frac{1}{K_0} \left(\partial_x \Theta_0 \right)^2 \right] \nonumber \\
&+& \int dx \; \frac{v_{1}}{2} \left[ K_{1} \left(\partial_x \Phi_{1} 
\right)^2 +  \frac{1}{K_{1}} \left(\partial_x \Theta_{1} \right)^2 \right]
\nonumber \\
&-& \int dx \; \left[  \frac{g_{\perp}}{a^2} 
\cos \left( \sqrt{2\pi} \Theta_{1} \right) 
+   \frac{g_{d}}{a^2} 
\cos \left( \sqrt{8\pi} \Phi_{1} \right) \right] .
\label{bosoN2bis}
\end{eqnarray}
There is a velocity anisotropy and two different Luttinger parameters which
stem from the marginal term of Eq. (\ref{bosoN2}) with coupling constant $\lambda$:
\begin{eqnarray}
K_{0} &=& \frac{K}{\sqrt{1+ \lambda v/K}}\nonumber \\
K_{1} &=& \frac{K}{\sqrt{1- \lambda v/K}} .
\label{Luttingerpara}
\end{eqnarray}
The effective Hamiltonian (\ref{bosoN2bis}) has been first studied in Ref. \onlinecite{orignacladder} in the context of
a bosonic two-leg ladder for incommensurate filling.
It  separates into two commuting pieces:
${\cal H}_{N=2} = {\cal H}_{0} + {\cal H}_{1}$ ($\left[{\cal H}_{0}, {\cal H}_{1} \right] = 0$).
The first contribution ${\cal H}_{0}$ of Eq. (\ref{bosoN2bis}) has the form of a Tomonaga-Luttinger Hamiltonian;
this signals that the bosonic field $\Phi_0$ is a gapless degree of freedom. All the non-trivial physics
is encoded in ${\cal H}_{1}$ which describes the competition between 
the intertube hopping and dipole-dipole interactions. 
As already discussed in Sec. II in the general $N$ case, there are two different phases which stem from this competition.
 When $1/2 < K_{1} <1$, ${\cal H}_{1}$ has a spectral gap $\Delta_{1}$ due  to 
 the cosine term with the bosonic field ${\Theta}_{1}$. As a result, 
 ${\Theta}_{1}$  is pinned onto one of the minima of 
 its sine-Gordon model, which we can choose to be $\langle  {\Theta}_{1}\rangle = {0}$ ($g_{\perp} > 0$). Thus, at low-energy $E \ll \Delta_{1}$, the creation operator of the bosons
simplifies as follows:
$ \Psi^{\dagger}_n  \sim \sqrt{\rho_0} \; e^{i \sqrt{\pi/2} \; \Theta_0}$, while
the $2 \pi \rho_0$ density operator is short-ranged since it depends on the 
strongly fluctuating field ${\Phi}_{1}$.
The resulting superfluid phase is then characterized by:
 \begin{eqnarray}
\langle  \Psi^{\dagger}_n \left(x \right) \Psi_m \left(0 \right) \rangle & \sim& x^{-1/(4K_{0})} 
\nonumber \\
\langle  \rho_n \left(x \right) \rho_m \left(0 \right) \rangle &  \sim& \rho^2_0 
- \frac{K_{0}}{4 \pi^2 x^2} ,
\label{superfluid2}
\end{eqnarray}
with $n,m = 1,2$.  In fact, in this phase, one may also consider the symmetric combination of the density with $4 \pi \rho_0$ 
oscillations in each tube: 
\begin{eqnarray}
	 \begin{array}{l}
	  \ds\rho^{4 \pi \rho_0} \sim \rho_0 \sum_{n=1}^{2} \cos
	\left(4 \pi \rho_0 x + 4 \sqrt{\pi}\varphi_{n}  \right)\\
	\qquad\;\;\sim 2 \rho_0 \cos
	\left(4 \pi \rho_0 x +  \sqrt{8\pi}  \Phi_0 \right) \cos
	\left(\sqrt{8\pi}  \Phi_{1} \right).
	 \label{cdw4kFdef}
	\end{array}
\end{eqnarray}
This operator is naively short-ranged in the phase with $\langle  {\Theta}_{1}\rangle = {0}$
since it depends on the field ${\Phi}_{1}$.
However, first-order perturbation theory with the Hamiltonian ${\cal H}_{1}$ cancels 
the contribution of $ \cos \left(\sqrt{8\pi}  \Phi_{1} \right)$ in $ \rho^{4 \pi \rho_0}$.
The low-energy description of that operator in the superfluid phase thus simplifies as follows:
 \begin{equation}
  \rho^{4 \pi \rho_0} \sim  2 \rho_0 \cos
\left(4 \pi \rho_0 x +  \sqrt{8\pi}  \Phi_0 \right).
 \label{cdw4kF}
\end{equation}
We deduce that the two-point correlation function of that operator has a power-law decay with exponent $4 K_0$.
Since $1/2<K_0<1$, the decay is much slower than the decays of the instabilities
of Eq. (\ref{superfluid2}).

\begin{figure}
     \begin{center}
     \includegraphics[width=0.99\columnwidth,clip]{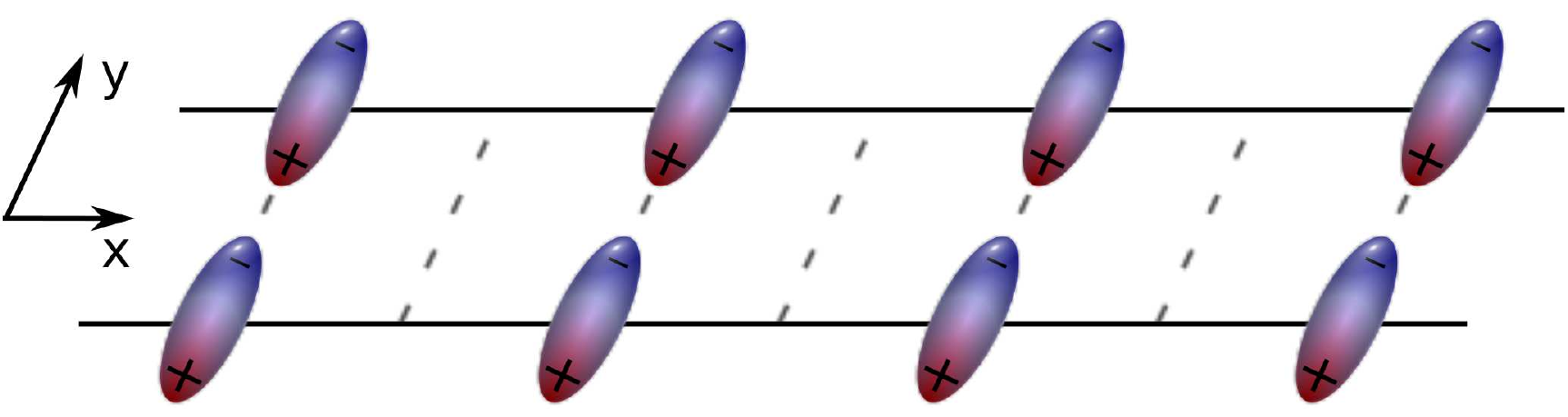}
\end{center}
\caption{(color online) The CDW order characterized by Eq. (\ref{cdw2}); the two CDW orders in each tube are in phase.}
\label{fig:CDW}
\end{figure}
The second phase is obtained
when $K_{1}<1/2$. 
The most relevant contribution of Eq. (\ref{bosoN2bis})
is now the cosine term with the bosonic field ${\Phi}_{1}$
which is pinned:
$\langle {\Phi}_{1} \rangle = 0$ ($g_d >0$). 
In this phase, the dual field ${ \Theta}_{1}$ is a strongly
fluctuating field. The resulting phase is a CDW phase where the two $2 \pi \rho_0$ CDW on each tube 
are in phase (see Fig. \ref{fig:CDW}):
\begin{eqnarray}
\langle  \Psi^{\dagger}_n \left(x \right) \Psi_m \left(0 \right) \rangle & \sim& e^{- x/\xi}
\nonumber \\
\langle  \rho_n \left(x \right) \rho_m \left(0 \right) \rangle &  \sim& \rho^2_0  + A \frac{\cos
\left(2 \pi \rho_0 x \right)}{x^{K_0}} ,
\label{cdw2}
\end{eqnarray}
$A$ being a non-universal constant and $\xi \sim v_{1}/\Delta_{1}$ is the correlation length which stems from 
the spectral gap of the Hamiltonian ${\cal H}_{1}$.
In this phase, the dimer instability $M^{\dagger} = \Psi^{\dagger}_1 \Psi^{\dagger}_2 \sim \rho_0 \; 
e^{i \sqrt{2\pi} \; \Theta_0}$ has 
a power-law correlation function with a larger decay $1/K_0$ ($K_0 < 1$).

The quantum phase transition between these two different phases occurs at 
$K_{1} = 1/2$. From Eq. (\ref{Luttingerpara}), we observe that  since $\lambda <0$, we have $K_{1} < K$, i.e., attractive intertube interactions such as in Eq. (\ref{dipoleinter2}) are favorable in order to reach the transition.
The effective Hamiltonian which governs this transition writes:
\begin{eqnarray}
&& {\cal H}_{\rm eff} = 
\int dx\;\frac{v_{1}}{2} \left[  \left(\partial_x \Phi_{1}\right)^2 
+\left(\partial_x \Theta_{1} \right)^2 \right]\qquad\qquad\quad\nonumber\\
&-&\int dx \; \left[  \frac{g_{\perp}}{a^2} 
\cos \left( \sqrt{4\pi} \Theta_{1} \right)+\frac{g_{d}}{a^2}\cos \left( \sqrt{4\pi} \Phi_{1} \right) \right].
\label{HameffN2}
\end{eqnarray}
This model is well known (see e.g. Ref. \onlinecite{ogilvie})
and can be exactly diagonalized by introducing two Majorana fermions $\xi^{1,2}$.
This procedure is nothing but the standard bosonization of two Ising
models \cite{zuber,boyanovskyising,bookboso}.
The refermionization rules are given by
\begin{eqnarray}
\xi_R^{1} + i \xi_R^{2} &=& \frac{1}{\sqrt{\pi a} } \exp\left(i\sqrt{4\pi}
\Phi_{1R}\right), \nonumber \\
\xi_L^{1} + i \xi_L^{2} &=& \frac{1}{\sqrt{ \pi a} } \exp\left(-i\sqrt{4\pi}
\Phi_{1L}\right),
\label{bosoising}
\end{eqnarray}
where the anticommutation between the right and left-moving Majorana
fermions is taken into account by the prescription: $\left[\Phi_{1R}, \Phi_{1L} \right] = i/4$.
The effective Hamiltonian can then be refermionized:
\begin{eqnarray}
&&{\cal H}_{\rm eff} = -\frac{i v_{1}}{2} \int dx \; \ \sum_{a=1}^{2}\left(
\xi_R^{a} \partial_x \xi_R^{a} -
\xi_L^{a} \partial_x \xi_L^{a} \right)
\nonumber \\
&-&  \int dx \;\left[ i m_1 \xi_R^{1} \xi_L^{1}
+ i m_2  \xi_R^{2} \xi_L^{2} \right],
\label{hamefffin}
\end{eqnarray}
where the two masses are given by:
\begin{eqnarray}
m_1 &=& \left(g_d - g_{\perp} \right)/a
\nonumber \\
m_2 &=& \left(g_d + g_{\perp} \right)/a .
\label{fermionsmasses}
\end{eqnarray}
The effective model (\ref{HameffN2}) at $K_{1} =1/2$ is thus
mapped onto two decoupled 1D Ising models in a transverse field.
The Majorana fermion $ \xi_{R,L}^{2}$ always has a positive mass $m_2$ which
means that, in our convention, the corresponding Ising model belongs to its
disordered phase where the disorder operator condenses: $\langle \mu_{2} \rangle \ne 0$.
In contrast, the Majorana fermion $ \xi_{R,L}^{1}$ has a smaller mass $m_1$ which can change
its sign; by tuning the hopping $g_{\perp}$, we are thus able to reach the Ising $\ZZ_2$ quantum critical point where $m_1 =0$.

The different order parameters defined in Eqs. (\ref{superfluid2}, \ref{cdw2}) can  be expressed in terms of the
Ising order ($\sigma_{1,2}$) and disorder
($\mu_{1,2}$) parameters of the two underlying Ising models.
In this respect, we need the Ising description of all the vertex operators with a scaling dimension
equal to $1/4$: \cite{zuber,boyanovskyising,shelton,allen,fabrizio}
\begin{eqnarray}
\sqrt{2} \cos(\sqrt{\pi} \Phi_{1}) 
&\sim&  \mu_1 \mu_2, \;  \sqrt{2} \cos(\sqrt{\pi} \Theta_{1}) 
\sim \sigma_1 \mu_2\quad\nonumber \\
\sqrt{2} \sin(\sqrt{\pi} \Phi_{1}) 
&\sim&  \sigma_1 \sigma_2, \; \sqrt{2} \sin(\sqrt{\pi} \Theta_{1}) 
\sim  \mu_1 \sigma_2 ,
\label{isingeries}
\end{eqnarray}
where the Ising order parameters are normalized at the $\ZZ_2$ quantum critical point
according to:
\begin{eqnarray}
\langle \sigma_a \left(\tau,x\right)  \sigma_b \left(0,0\right) \rangle
= \frac{\delta_{ab}}{\left(v_{1}^2 \tau^2 + x^2 \right)^{1/8}},
\label{isingcorrelation}
\end{eqnarray}
with $a,b = 1,2$, and similarly for the two-point correlations of the Ising disorder
operators. 
At low-energy, i.e., $E \ll m_2$, we can average the operators of the second Ising model
($\langle \mu_{2} \rangle \ne 0$ and $\langle \sigma_{2} \rangle = 0$).
The bosonic creation and density operators then simplify as follows, using Eq. (\ref{isingeries}):
\begin{equation}
\begin{array}{lll}
	\Psi^\dagger_n &\sim&\ds e^{i\sqrt{\pi/2K_0}\;\Theta_0}\sigma_1\\
	\rho_n &\sim&\ds\rho_0+\sqrt{\frac{K_0}{2\pi}}\partial_x \Phi_0\\
	&&\ds\qquad\quad+\sqrt{A}\cos\left[2\pi\rho_0 x + \sqrt{2\pi K_0}\Phi_0\right]\mu_1,
\end{array}\label{descript}\end{equation}
with $K_0 \simeq 1/2$.
In particular, we observe that the  bosonic creation operator expresses in terms
of the Ising order operator $\sigma_1$, in full agreement with the identification
of the discrete $\ZZ_N$ symmetry (\ref{Zsym}) made in Sec. II. 

It is then straightforward to deduce the zero-temperature phase diagram of model (\ref{HameffN2}). When $g_{\perp} < g_d$, i.e. $m_1 >0$, the Ising model belongs to its disorder phase
with $\langle \sigma_{1} \rangle = 0$ and $\langle \mu_{1} \rangle \ne 0$ . We recover the CDW phase with the leading asymptotics (\ref{cdw2}). When $g_{\perp} > g_d$,  the Ising model belongs to its ordered phase with $\langle \sigma_{1} \rangle \ne 0$ and the double-tube system sits in its superfluid phase (\ref{superfluid2}). The quantum phase transition between these two phases occurs for $g_{\perp} = g_d$
and is driven by the Ising degrees of freedom. The resulting critical properties are determined by the low-energy description (\ref{descript}) and correlations (\ref{isingcorrelation}):
\begin{equation}\begin{array}{lll}
	\langle\Psi^\dagger_n\left(x\right)\Psi_m \left(0\right)\rangle
	&\sim&\ds x^{-1/(4K_0)-1/4}\\
	\langle\rho_n\left(x\right)\rho_m\left(0\right)\rangle
	&\sim&\ds \rho_0^2-\frac{K_0}{4\pi^2 x^2}+A\frac{\cos\left(2\pi\rho_0 x\right)}{x^{K_0+1/4}},
\end{array}\label{transition2}\end{equation}
with $K_0 \simeq 1/2$.
The quantum phase transition is thus described by a U(1) $\times$  $\ZZ_2$ CFT
with a central charge $c= 1 + 1/2= 3/2$. The SU(2)$_2$ critical properties of
the perturbation (\ref{sdgint}) for $N=2$ 
are weakly broken down to U(1) $\times$  $\ZZ_2$ since the bosonic fields
$ \Phi_0$ and $\Phi_{1}$  have different Luttinger parameters (\ref{Luttingerpara})
and velocities.

\section{Triple-tube systems}

\begin{figure}[ht]
     \begin{center}
     \includegraphics[width=0.99\columnwidth,clip]{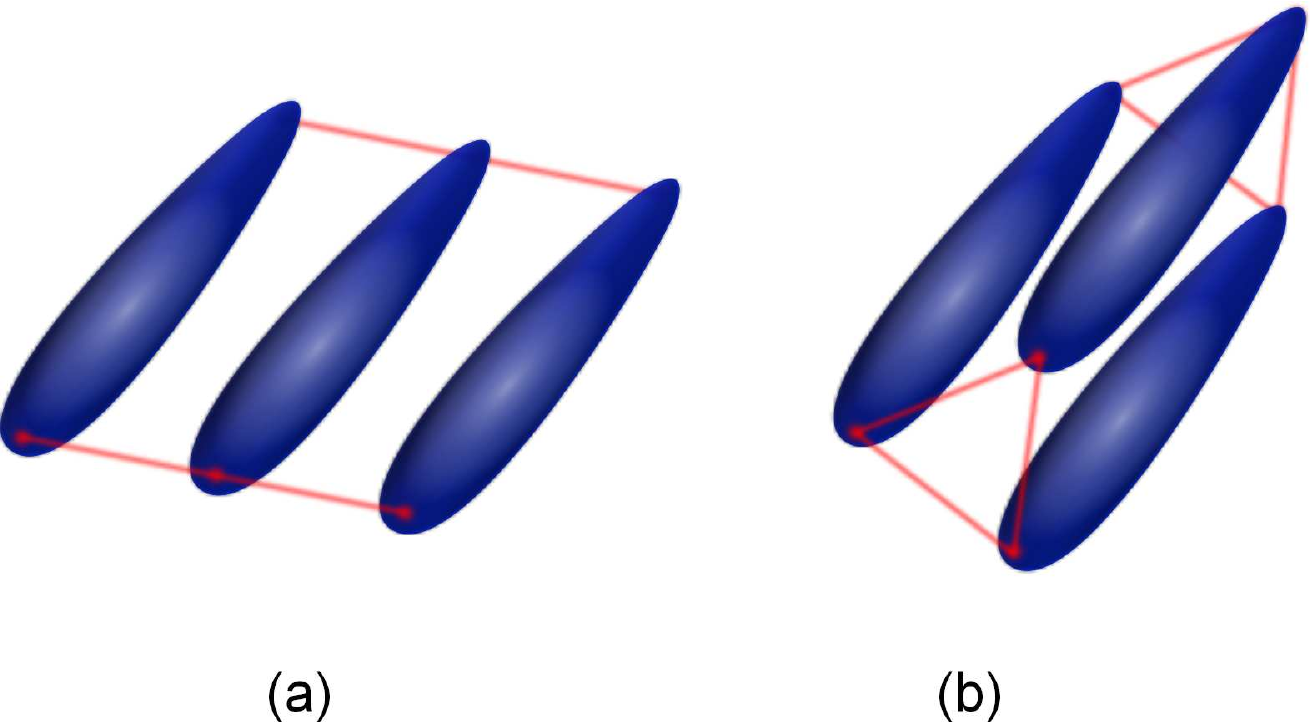}
\end{center}
\caption{A triple-tube system for (a) open boundary conditions; (b) periodic boundary conditions. As in Fig. \ref{fig:bosonsN}, the tubes are separated by a distance $d$.}
\label{fig:bosonsN3}
\end{figure}

In this section, we determine the zero-temperature physical properties of the triple-tube system
 which corresponds to model (\ref{ham}) with $N=3$ (see Fig. \ref{fig:bosonsN3}).
 We consider here two different geometries: one with open-boundary conditions (Fig. \ref{fig:bosonsN3}(a))
 and the other with periodic boundary conditions, where the tubes form an equilateral triangle (see Fig. \ref{fig:bosonsN3}(b)).
 In the latter case, we assume a non-frustrated situation where 
 all intertube density-density interactions are attractive.

\subsection{$\ZZ_3$  emerging quantum criticality}
The harmonic-fluid representation for the triple-tube system of  Fig. \ref{fig:bosonsN3}(a) is 
given by model (\ref{hamluttinger}, \ref{bosointer}) with $N=3$.
Here, we follow the general strategy of Sec. II and first consider the most
relevant contribution, i.e. Eq. (\ref{hameff}) with $N=3$.
We single out the gapless U(1) $\Phi_0$ bosonic field by switching to the basis (\ref{SUNbasis}) with $N=3$. The inverse transformation (\ref{invtrans}) explicitly reads as follows:
\begin{eqnarray}
\varphi_1 &=& \frac{1}{\sqrt 3} \Phi_0 +  \frac{1}{\sqrt 2} \Phi_1 +  \frac{1}{\sqrt 6} \Phi_2\nonumber\\
\varphi_2 &=& \frac{1}{\sqrt 3} \Phi_0 -  \frac{2}{\sqrt 6} \Phi_2\nonumber\\
\varphi_3 &=& \frac{1}{\sqrt 3} \Phi_0 -  \frac{1}{\sqrt 2} \Phi_1 +  \frac{1}{\sqrt 6} \Phi_2,
\label{invtransN3}
\end{eqnarray}
with similar definitions for the dual fields $\vartheta_{1,2,3}$.
The effective Hamiltonian (\ref{hameff}) for $N=3$ separates into two commuting pieces:
${\cal H}_{\rm eff} = {\cal H}_{0} + {\cal H}_{s}$, where ${\cal H}_{0}$ takes
the form of the Tomonaga-Luttinger model:
\begin{equation}
{\cal H}_{0} =  \int dx \; \frac{v}{2} \left[ K \left(\partial_x \Phi_0
\right)^2 +  \frac{1}{K} \left(\partial_x \varTheta_0 \right)^2 \right].
\label{hamluttinger0}
\end{equation}	 
The remaining degrees of freedom in ${\cal H}_{s}$ are strongly coupled and control
the quantum phase transition between the superfluid phase (\ref{superfluid}) and 
the CDW phase (\ref{cdw}) at $K=1/2$.  When $K=1/2$ and in the new basis, the effective Hamiltonian for these degrees of freedom is given by:
\begin{eqnarray}
&& {\cal H}_{s} =  \int dx \; \frac{v}{2} \left[ \left(\partial_x \Phi_1 
\right)^2 +  \left(\partial_x \Theta_1 \right)^2 + 1 \rightarrow 2 \right] \nonumber \\
&-&  \int dx \; \left[   \frac{2 g_{\perp}}{a^2}
\cos \left( \sqrt{\pi} \; \Theta_{1} \right)   \cos \left( \sqrt{3\pi} \; \Theta_{2} \right) 
\right.
\nonumber \\
&+&  \left. \frac{2 g_{d}}{a^2} \cos \left( \sqrt{\pi} \Phi_{1}  \right)  \cos \left( \sqrt{3\pi} \Phi_{2} \right) \right] .
\label{bosoN3}
\end{eqnarray}
The bosonic field $\Phi_{1}$ is at the free-fermion point, i.e. the terms
of the form $\cos \left( \beta \Phi_{1}  \right)$ or 
$\cos\left( \beta \; \Theta_{1} \right)$ that appear in the Hamiltonian 
all have $\beta=\sqrt{\pi}$. For this reason, we can introduce two effective Ising models
with order parameters $\sigma_{1,2}$ and disorder fields $\mu_{1,2}$, 
so as to refermionize the vertex operators of Eq. (\ref{bosoN3}) with a scaling dimension $1/4$, just
as we did for the double-tube system (see Eq. (\ref{isingeries})):
\begin{eqnarray}
&& {\cal H}_{s} = -\frac{i v}{2} \int dx \; \ \sum_{a=1}^{2}\left(
\xi_R^{a} \partial_x \xi_R^{a} -
\xi_L^{a} \partial_x \xi_L^{a} \right)
\nonumber \\
 &+& \int dx \; \frac{v}{2} \left[ \left(\partial_x \Phi_2 
\right)^2 +  \left(\partial_x \Theta_2 \right)^2 \right] \nonumber \\
&-& \frac{\sqrt{2}}{a}  \int dx \; \left[   g_{\perp} 
\sigma_1 \mu_2  :\cos \left( \sqrt{3\pi} \; \Theta_{2} \right): 
\right.
\nonumber \\
&+& \left. g_{d}  \mu_1 \mu_2  :\cos \left( \sqrt{3\pi} \Phi_{2} \right): \right] .
\label{bosoN3bis}
\end{eqnarray}

The next step of the approach is to switch to
a different basis to determine the main
effect of the relevant perturbation of
model (\ref{bosoN3bis}).  In this respect, we exploit the fact that
the CFT defined by the  product of one Ising CFT (the one with disorder 
operator $\mu_1$) and the CFT associated to the boson $\Phi_2$, can be described in terms of the product ${\cal M}_4 \times {\cal M}_5$ CFT, i.e.,  TIM $\times$ $\ZZ_3$  Potts CFTs. 
Indeed, the TIM and the $\ZZ_3$ Potts CFTs have central charges $c=7/10$ and $c=4/5$ respectively, so that the sum gives $c=3/2$, the total central charge of an Ising CFT $c=1/2$ and a CFT associated with one boson $c=1$.
The TIM CFT describes the physical properties of the two-dimensional dilute Ising model
at its tricritical point \cite{dms,mussardo}.
The precise identification of the conformal embedding has been derived
by the authors of Ref. \onlinecite{bul} and requires a projection 
which  restricts the primary operators of ${\cal M}_4 \times {\cal M}_5$
to the subset: $\{\Phi^{(4)}_{r,s} \Phi^{(5)}_{s,q} \}$ where 
$\Phi^{(p)}_{r,s}$ ($ 1 \le r \le p-1, 1 \le s \le p$)  denotes the primary fields of the 
minimal model series ${\cal M}_p$ CFT with scaling
dimensions $\Delta^{(p)}_{r,s} =  \frac{\left[\left(p+1\right)r - p s \right]^2 - 1}{2 p \left(p+1\right)}$
\cite{dms}.
In Sec. 3 of the Appendix, we show that we have the following correspondences:
\begin{eqnarray}
\mu_1 :\cos\left(\sqrt{3 \pi} \Phi_2\right):  +
\sigma_1 :\cos\left(\sqrt{3 \pi} \Theta_2\right): &\sim& \sigma_{\rm TIM}^{'}\label{CFTidentities}\\
\mu_1 :\cos\left(\sqrt{3 \pi} \Phi_2\right):  -
\sigma_1 :\cos\left(\sqrt{3 \pi} \Theta_2\right): &\sim& \sigma_{\rm TIM} \epsilon_{\ZZ_3},\nonumber
\end{eqnarray}
$\sigma_{\rm TIM}$ and $\sigma_{\rm TIM}^{'}$  being respectively the magnetization
and subleading magnetization operators of the TIM CFT 
with scaling dimensions $3/40$ and $7/8$ \cite{dms}.
In Eq. (\ref{CFTidentities}), $\epsilon_{\ZZ_3}$ refers to the thermal operator 
of the $\ZZ_3$  Potts CFT with scaling dimension $4/5$.

With all these results, one can now express the low-energy effective Hamiltonian
in terms of these new degrees of freedom:
 \begin{eqnarray}
&& {\cal H}_{s} = -\frac{i v}{2} \int dx \; \left(
\xi_R^{2} \partial_x \xi_R^{2} -
\xi_L^{2} \partial_x \xi_L^{2} \right)  
+{\cal H}^{{\rm TIM}}_{0} 
+ {\cal H}^{\ZZ_3}_{0}
 \nonumber \\
&-& \frac{\sqrt{2}}{2a}  \int dx \;  \left[   \left(g_{\perp} + g_{d}  \right) \mu_2 \sigma_{\rm TIM}^{'}
\right.  \nonumber \\
&+& \left. 
\left(- g_{\perp} + g_{d}  \right) \mu_2 \sigma_{\rm TIM} \epsilon_{\ZZ_3}  \right],
\label{bosoN3fin}
\end{eqnarray}
where $ {\cal H}^{{\rm TIM}}_{0}$ and ${\cal H}^{\ZZ_3}_{0}$ respectively denote the
Hamiltonians of the TIM and $\ZZ_3$ CFTs.
The operator $\mu_2 \sigma_{\rm TIM}^{'}$ is a strongly relevant perturbation with scaling dimension 1 which couples
the Ising model associated to the Majorana fermion $\xi^{2}$ with the TIM CFT.
The Ising critical point is destabilized upon switching on the disorder
field $\mu_2$, while the $\sigma_{\rm TIM}^{'}$ perturbation is known 
to be a massive integrable deformation of the TIM CFT
for all signs of its coupling constant \cite{mussardo}. We deduce that a spectral gap $\Delta$
for the Ising and TIM degrees of freedom is opened as soon as the tubes are coupled.
In the low-energy limit where $E \ll \Delta$, the low-energy effective Hamiltonian (\ref{bosoN3fin}) 
then simplifies as
\begin{eqnarray}
 {\cal H}_{s} = {\cal H}^{\ZZ_3}_{0}
+  A \int dx \;  \left(- g_{\perp} + g_{d}  \right) \epsilon_{\ZZ_3}  ,
\label{Pottsdeform}
\end{eqnarray}
$A$ being a non-universal constant. This model is a massive and integrable
perturbation of the $\ZZ_3$ CFT \cite{fateev}. A spectral gap is thus generated, except at
the fine-tuned point $g_{\perp} =  g_{d} $ where the model displays a quantum critical
behavior in the  $\ZZ_3$ Potts universality  class. The position of the critical point, 
i.e. $g_{\perp} =  g_{d} $, corresponds to the self-duality symmetry of model (\ref{bosoN3}):
$\Phi_{1,2} \leftrightarrow \Theta_{1,2}$.

Adding the contribution of the gapless boson $\Phi_0$, we deduce that the 
quantum phase transition of the triple-tube system is governed by the SU(2)$_3$ CFT
with central charge $c= 1 + 4/5 = 9/5$; this is in full agreement with the general analysis of Sec. II.

As in the $N=2$ case, the marginal forward-scattering operator ${\cal O}_{\text fs} = \partial_x \varphi_{2} 
(\partial_x \varphi_{1} +  \partial_x \varphi_{3})$ is expected to introduce anisotropies and the symmetry
of the quantum critical point is U(1) $\times$ $\ZZ_3$.
In particular, using the new basis (\ref{invtransN3}), we find that
\begin{equation}
{\cal O}_{\text fs} = \frac{2}{3} \left[ (\partial_x \Phi_{0})^2 -  (\partial_x \Phi_{2})^2 \right]
-   \frac{2}{3 \sqrt{2}} \partial_x \Phi_{0} \partial_x \Phi_{2} .
\label{luttN3}
\end{equation}
As expected, the first term of this equation merely introduces velocity anisotropies and different Luttinger parameters
for the bosons $ \Phi_{0,1,2}$ as in the double-tube case. On the other hand, the last term gives
a residual coupling between the gapless boson $\Phi_{0}$ and the remaining degrees of freedom.
Such term has also be encountered in the analysis of the zero-temperature phase diagram of 1D three-component
cold fermions; its main effect is not clear \cite{azaria}. However, we think that the latter perturbation
will not destroy the quantum criticality of the phase transition of the triple-tube model.

In any case, one can eliminate this term by considering a triple-tube system with transverse periodic boundary
conditions as depicted in Fig. \ref{fig:bosonsN3}(b). In that case, the  forward-scattering perturbation 
of this problem is:
\begin{eqnarray}
{\cal O}^{\Delta}_{\text fs} &=&\partial_x \varphi_{2} 
(\partial_x \varphi_{1} +  \partial_x \varphi_{3}) + \partial_x \varphi_{1} \partial_x \varphi_{3} 
\nonumber \\
&=& (\partial_x \Phi_{0})^2 - \frac{1}{2} \left[ (\partial_x \Phi_{1})^2 + (\partial_x \Phi_{2})^2 \right],
\label{luttN3P}
\end{eqnarray}
 so that this contribution is exhausted by velocities and Luttinger parameters renormalization.	
At the self-dual point  $g_{\perp} =  g_{d}$, the effective Hamiltonian which controls
the transition of the model has a similar structure as in the open boundary case (see Eq. (\ref{bosoN3fin})):
  \begin{eqnarray}
 {\cal H}^{\Delta}_{s} &=& -\frac{i v}{2} \int dx \; \left(
\xi_R^{2} \partial_x \xi_R^{2} -
\xi_L^{2} \partial_x \xi_L^{2} \right)  
+{\cal H}^{{\rm TIM}}_{0} 
+ {\cal H}^{\ZZ_3}_{0}
 \nonumber \\
&-& \frac{\sqrt{2} g_{d} }{a}  \int dx \;   \mu_2 \sigma_{\rm TIM}^{'}
-  \frac{2 \pi g_{d} }{a}  \int dx \;  i \xi_R^{2} \xi_L^{2} .
\label{TransitionTripleP}
\end{eqnarray}
We observe that here, the spectral gap for the Ising and TIM degrees of freedom is now explicit,
due to the mass term ($m_2 = 2 \pi g_{d} /a$) for the 
Majorana fermion $\xi^{2}$. Indeed, since this mass is positive, the corresponding
Ising model belongs to its disordered phase with $\langle \mu_2 \rangle \ne 0$. Averaging out these degrees
of freedom, one is left with a TIM CFT perturbed by the  subleading magnetization operator $\sigma_{\rm TIM}^{'}$, 
which gives a spectral gap \cite{mussardo}. 
We conclude, as in the open case, that at low-energy, the $\ZZ_3$ degrees of freedom
remain untouched and display quantum criticality with central charge $c=4/5$.

Our approach allows us to determine the leading asymptotics of the correlation functions at the 
quantum critical point with U(1) $\times$ $\ZZ_3$ symmetry.
The bosonic creation operators $\Psi^{\dagger}_n$ can be expressed in terms of
the bosons (\ref{invtransN3}) with $K=1/2$: 
 \begin{eqnarray}
 \Psi^{\dagger}_{1,3} &\sim& e^{i \sqrt{ \pi/3K_0} \; \Theta_0} e^{\pm i \sqrt{\pi} \; \Theta_1}
 e^{i \sqrt{\pi/3} \; \Theta_2} 
  \nonumber \\
  \Psi^{\dagger}_{2} &\sim& e^{i \sqrt{\pi/3K_0} \; \Theta_0} e^{i \sqrt{4\pi/3} \; \Theta_2} ,
 \label{psin}
\end{eqnarray}
with $K_0 \simeq 1/2$.
Using Eq. (\ref{isingeries}) and taking into account that the Ising model with disorder parameter
$\mu_2$ sits in its disordered phase, we have
\begin{eqnarray}
 \Psi^{\dagger}_{1,3} &\sim& e^{i \sqrt{ \pi/3K_0} \; \Theta_0} \sigma_1
 e^{i \sqrt{\pi/3} \; \Theta_2} 
  \nonumber \\
  \Psi^{\dagger}_{2} &\sim& e^{i \sqrt{\pi/3K_0} \; \Theta_0} e^{i \sqrt{4\pi/3} \; \Theta_2} .
 \label{psinbis}
\end{eqnarray}
The next step is to express the operators $\sigma_1 e^{i \sqrt{\pi/3} \; \Theta_2}$ and
$ e^{i \sqrt{4\pi/3} \; \Theta_2}$, with respectively scaling dimensions $5/24$ and $1/3$,
in terms of the TIM $\times$ $\ZZ_3$ degrees of freedom. We find the following identification:
\begin{eqnarray}
 \sigma_1 e^{i \sqrt{\pi/3} \; \Theta_2} 
 &\sim& \sigma_{\rm TIM} \sigma_{\ZZ_3}
   \nonumber \\
  e^{i \sqrt{4\pi/3} \; \Theta_2} &\sim& \epsilon_{\rm TIM} \sigma_{\ZZ_3},
 \label{TIMZ3indent}
\end{eqnarray}
$\sigma_{\ZZ_3}$ being the $\ZZ_3$ spin operator with scaling dimension $2/15$, and
$\epsilon_{\rm TIM}$, the thermal operator of the TIM CFT with scaling dimension $1/5$.
Finally, at low-energy, i.e. for energies smaller than the spectral gap of the TIM degrees of freedom,
the  bosonic creation operators $\Psi^{\dagger}_n$  of the triple-tube system simplify as follows:
\begin{eqnarray}
\Psi^{\dagger}_{n} &\sim& e^{i \sqrt{\pi/3K_0} \; \Theta_0} \sigma_{\ZZ_3} .
 \label{IRpsin}
\end{eqnarray}
We observe that $\Psi^{\dagger}_{n}$ expresses directly in terms the $\ZZ_3$ spin operator 
in full agreement with the identification (\ref{Zsym}) in the general case. 
Similarly, we can find the low-energy representation of the $2 \pi \rho_0$ CDW on each tube
$ \rho^{2\pi \rho_0}_n$ at $K=1/2$:
\begin{eqnarray}
\rho^{2\pi \rho_0}_n &\sim& e^{i 2\pi \rho_0 x + \sqrt{4 \pi K} \; \varphi_n}
 \nonumber \\
&\sim& e^{i 2\pi \rho_0 x + i \sqrt{4 \pi K_0/3} \; \Phi_0} \mu_{\ZZ_3} ,
 \label{IRdens}
\end{eqnarray}
with $K_0 \simeq 1/2$ and $ \mu_{\ZZ_3}$  is the $\ZZ_3$ disorder spin operator which is
dual to the $\ZZ_3$ spin operator $\sigma_{\ZZ_3}$.
Using the results (\ref{IRpsin}, \ref{IRdens}), we deduce the leading asymptotics
of the equal-time correlations at the quantum critical point of the triple-tube model: 
\begin{eqnarray}
\langle  \Psi^{\dagger}_n \left(x \right) \Psi_m \left(0 \right) \rangle & \sim& x^{-1/(6K_{0})-4/15} 
\nonumber \\
\langle  \rho_n \left(x \right) \rho_m \left(0 \right) \rangle &  \sim& \rho^2_0 
- \frac{K_{0}}{6 \pi^2 x^2} + A \frac{\cos
\left(2 \pi \rho_0 x \right)}{x^{2K_0/3+4/15}} ,
\label{transition3}
\end{eqnarray}
with $K_0 \simeq 1/2$.
The quantum phase transition is thus described by a U(1) $\times$  $\ZZ_3$ CFT
with central charge $c= 1 + 4/5= 9/5$. 

\subsection{Stability of the quantum critical point and the $\ZZ_3$  chiral clock model}

The next step of the approach is to investigate the stability of this quantum critical point 
to generic perturbations allowed by the symmetries of the model.
This includes all irrelevant perturbations, that we have neglected in the continuum description, 
which may change their status at the new U(1) $\times$  $\ZZ_3$
fixed point.
The entire content of the $\ZZ_3$ Potts model
is known\cite{dms} and the thermal operator $\epsilon_{\ZZ_3}$ is the 
only relevant operator having zero
conformal spin and preserving the global $\ZZ_3$ symmetry of the
model.
 The presence of two independent coupling constants in the model, namely
$J_{\perp}$ and $C_{dd}$, should be enough to kill this relevant operator and reach the $\ZZ_3$  critical point.
A similar criticality has been found in 1D XXZ Heisenberg chain in magnetic fields \cite{phleorignac}.
The other relevant operators of the $\ZZ_3$  Potts model
carry a non-zero conformal spin: $\Phi_{(2/5,7/5)}$ and $\Phi_{(7/5,2/5)}$
are $\ZZ_3$  operators with scaling dimension $9/5$ and conformal spin $ S = \pm 1$.
These non-Lorentz invariant perturbations, if generated, might drive the system to the fixed point 
of the chiral three-state Potts universality class \cite{cardychir}; the general low-energy effective field theory of triple-tube systems, which controls the vicinity
of the $\ZZ_3$  fixed point, would then read as follows:
\begin{equation}
{\cal H}^{\ZZ_3}_{\delta} = {\cal H}^{\ZZ_3}_{0}
+ \int dx \;  \delta  \Phi_{(2/5,7/5)}
+ \delta^{*}  \Phi_{(7/5,2/5)} 
+   \delta_g  \;\epsilon_{\ZZ_3} ,
\label{chireff}
\end{equation}
where $\delta$ and $\delta_g = A \left(- g_{\perp} + g_{d}  \right)$ are small coupling constants describing
the departure from the $\ZZ_3$  quantum critical point.
It is thus important to find out if the relevant non-scalar operators $\Phi_{(2/5,7/5)}, \Phi_{(7/5,2/5)}$
are permitted in the triple-tube systems investigated here. 

In this respect, we observe that a non-zero conformal spin perturbation emerges in the continuum limit of
the triple-tube system with open boundary conditions (Fig. 3(a)), when expanding the square root in the harmonic-fluid representation of the bosonic operator (\ref{bosoop}); for $K=1/2$, it reads:
\begin{equation}
{\cal O}_{S=\pm 1} = \sum_{n=1}^{2}  \left( \partial_x \varphi_{n+1} + \partial_x \varphi_{n} \right)
\cos \left( \sqrt{2\pi} \left( \vartheta_{n+1} -  \vartheta_{n} \right) \right).
\label{twist}
\end{equation}
This term is a marginal perturbation with conformal spin $ S = \pm 1$, and it is invariant under parity,
since $x\rightarrow - x$ , $\varphi_{n}(x) \rightarrow - \varphi_{n}(-x)$, which is indeed a symmetry
of Eqs. (\ref{densboso},\ref{bosoop}). 
In the triple-tube model with periodic boundary conditions (Fig. 3(b)), the non-Lorentz invariant perturbation becomes:
\begin{equation}
{\cal O}^{\Delta}_{S=\pm 1} = \sum_{n=1}^{3}  \left( \partial_x \varphi_{n+1} + \partial_x \varphi_{n} \right)
\cos \left( \sqrt{2\pi} \left( \vartheta_{n+1} -  \vartheta_{n} \right) \right).
\label{twistperiodic}
\end{equation}
Using the basis (\ref{invtransN3}) and the refermionization (\ref{bosoising}, \ref{isingeries}), we focus on the following combinations for each triple-tube system:
\begin{eqnarray}
{\cal O}^{(1)}_{S=\pm 1} &=& \frac{1}{\sqrt{2}} \mu_2 \sigma_1 : \partial_x \Phi_2 \cos \left( \sqrt{3\pi} \Theta_{2} \right):
 \nonumber \\
{\cal O}^{\Delta}_{S=\pm 1} &=& {\cal O}^{(1)}_{S=\pm 1} + \partial_x \Phi_2  i \pi \xi_R^{1} \xi_L^{1},
\label{twistpert}
\end{eqnarray}
where the second operator occurs for the triple-tube model with periodic boundary conditions.
We can now express these non-Lorentz invariant perturbations in terms of 
the TIM $\times$ $\ZZ_3$ degrees of freedom. We find the following identification up to
some unimportant normalization factors:
\begin{eqnarray}
{\cal O}^{(1)}_{S=\pm 1} &\sim& \mu_2 \sigma_{\rm TIM} \left[ \Phi_{(2/5,7/5)}
+ \Phi_{(7/5,2/5)} \right] 
 \nonumber \\
 \partial_x \Phi_2  i \xi_R^{1} \xi_L^{1}&\sim& \epsilon_{\rm TIM} \left[ \Phi_{(2/5,7/5)}
+ \Phi_{(7/5,2/5)} \right] .
\label{twistpertfin}
\end{eqnarray}
Thus, at energies much smaller than the spectral gap of the Ising and TIM degrees of freedom, the 
physical properties of triple-tube systems are governed by the effective Hamiltonian
(\ref{chireff}) with $\delta  = \delta^{*}$ (and $  \delta_g  = A (- g_{\perp} + g_{d})$).
The latter Hamiltonian is known to describe the physical properties of the 2D 
uniaxial  $\ZZ_3$ chiral Potts model, the so-called Ostlund-Huse model \cite{ostlundhuse},
 in the vicinity of the three-state Potts critical point \cite{cardychir}. Interestingly enough, the competition between superfluidity and CDW in triple-tube systems is thus similar to the domain-wall wetting
transition in chiral clock model \cite{ostlundhuse,husefisher}.
We note that a 1D boson chain with constraints also has physical properties described 
by the  $\ZZ_3$ chiral Potts model \cite{fendley}.
\begin{figure}
     \begin{center}
     \includegraphics[width=0.85\columnwidth,clip]{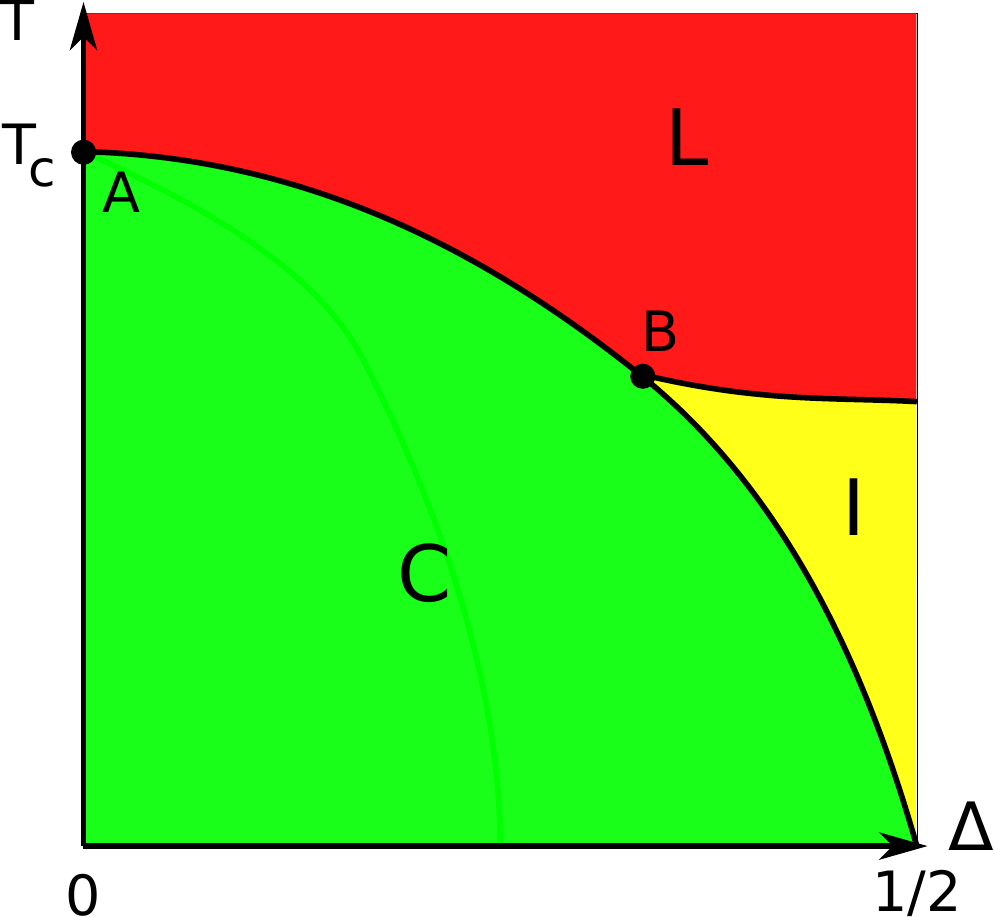}
\end{center}
\caption{(color online) Schematic phase diagram of the 2D $\ZZ_3$ chiral clock model; 
C, L, and I denote respectively the commensurate, liquid, and incommensurate phase;
$T_c$ is the critical temperature of the three-state Potts model and $\Delta$
the asymmetry parameter.}
\label{fig:PhaseDiagPotts}
\end{figure}

The Ostlund-Huse model on the square lattice is defined as follows:
\begin{eqnarray}
{\cal H}_{\rm OH} &=& - J \sum_{<i,j>} \cos \left[ \frac{2 \pi}{3} \left(n_i - n_j + \Delta\right) \right]
 \nonumber \\
&-& J \sum_{<i,k>} \cos \left[ \frac{2 \pi}{3} \left(n_i - n_k \right) \right],
\label{OHmodel}
\end{eqnarray}
where a variable $n_i = 0,1,2$ is associated with each site $i$ of a square lattice
and the sum $<i,j>$ (respectively $<i,k>$) 
are taken over nearest neighbours  in the axial, e.g. $x$, (respectively $y$) direction. 
The Ostlund-Huse model displays a global $\ZZ_3$ symmetry and has been introduced to describe the 
commensurate-incommensurate transition observed in monolayers absorbed on rectangular substrates \cite{ostlundhuse}.
For $\Delta =0$, the model reduces to the ordinary  2D $\ZZ_3$ clock model, 
which is equivalent to the three-state Potts model, and enjoys a S$_3$ (the permutation
group of three objects) global symmetry.
Such a model displays a second-order phase transition from a commensurate (or ferromagnetic) phase
to a liquid (or paramagnetic) phase in the $\ZZ_3$ parafermionic universality class.
The phase diagram of the model with uniaxial anisotropy $\Delta \ne 0$ has proved complicated and very controversial
(see Refs. \onlinecite{dennijs,perk} for a review).
The main effect of the asymmetry parameter $\Delta$ is to introduce a new incommensurate 
gapless phase (the so-called floating phase, with central charge $c=1$) between the 
conventional commensurate and liquid phases (see Fig. 4 for a schematic phase diagram).
The floating phase is characterized by a modulated order with a wavevector which varies continuously
with $T$ and $\Delta$. There is considerable controversy over the possible existence
of the Lifshitz point (point B in Fig. 4). To the best of our knowledge, the situation is still not settled.
Three different scenari are possible:

(a) There is a Lifshitz point at finite value of $\Delta$ and the Potts critical point (point A in Fig. 4)
extends along the AB line of Fig. 4 \cite{ostlundhuse, kadanoff, duxbury}.

(b) The Lifshitz point moves to $\Delta=0$: there is no direct transition between the commensurate
and liquid phase but an intermediate floating phase  \cite{haldanePotts,schulzPotts,rittenberg}.
There are two phase transitions to reach the liquid
phase from the commensurate one: first, a commensurate-incommensurate transition followed by a BKT
transition.

(c)  There is a Lifshitz point at finite value of $\Delta$ but the AB line of Fig. 4 is a new critical line with
critical exponents different from the three-state Potts universality class \cite{husefisher,rittenbergbis,sato}.

In our triple-tube systems, the case (a) corresponds to a U(1) $ \times \ZZ_3$ quantum phase transition
between the CDW and superfluid phases, described by the properties (\ref{transition3}).
In the second case (b), there is no direct transition between these two phases but 
an intermediate incommensurate gapless phase will then emerge.
Adding the contribution of the gapless boson $\Phi_0$, this incommensurate gapless
phase has central charge $c=1+1=2$. The correlation functions (\ref{transition3}), in this floating phase, will be non-universal
and incommensurate with a wave-vector $Q \ne 2 \pi \rho_0$. We expect the extent of this incommensurate
phase to be tiny since $\delta  = \delta^{*}$ is small in our approach.
Finally, in the last case (c), we still have a direct quantum phase transition between the 
CDW and superfluid phases. The correlations functions, at the quantum critical point,
are still described by Eqs. (\ref{transition3}) but with different (unknown) universal exponents.

Unfortunately, the low-energy properties of the effective field theory (\ref{chireff}) with $\delta  = \delta^{*}$ and 
$\delta_g  = 0$ are not known and cannot help to resolve the controversy \cite{cardychir}. However, on general grounds, 
a relevant $S = \pm 1$ conformal spin perturbation
usually gives rise to some incommensurate behavior \cite{cardychir,nersesyantwist,allenphle,tsveliktwist}.
The second case, i.e. the existence of an intermediate incommensurate gapless phase, thus seems
likely. However, relatively recent density-matrix-renormalization group calculations on 
the Ostlund-Huse model (\ref{OHmodel}) support scenario (c) and no incommensurate gapless phase
is found when $\Delta < 1/4$ \cite{sato}.
In addition, one might expect a different behavior between the two triple-tube systems of Fig. 3
since they do not share the same symmetry. In the equilateral triangle situation, we have a lattice global
S$_3$ invariance in contrast to the open case (Fig. 3(a)). Since the Ostlund-Huse model 
does not have this invariance except at $\Delta =0$, it is tempting to conjecture that $\delta  = \delta^{*} =0$
in the low-energy effective Hamiltonian (\ref{chireff}) for the periodic case. In that case,
the quantum phase transition between the superfluid and CDW phases belongs to the U(1) $ \times \ZZ_3$ 
universality class.
Unfortunately, within our
approach, we cannot directly check that $\delta  = \delta^{*} =0$. Indeed, the evaluation of 
these coupling constants requires the exact knowledge
of the non-universal amplitudes on Ising and TIM degrees of freedom in Eq. (\ref{twistpertfin})
for the theory (\ref{TransitionTripleP}).
In this respect, it would be very interesting in the future,
to numerically investigate the zero-temperature phase diagram of 
lattice triple-tube models of Fig. 3. In particular, it would further shed light on the nature of the 
quantum phase transition and its relation to the physics of the 2D $\ZZ_3$ chiral Potts model.

\section{Concluding remarks}

In summary, we have investigated the low-energy properties of a system of $N$
 1D dipolar bosons tubes coupled together via intertube hopping and dipole-dipole
 interactions. 
 Here, we have focused our analysis on the quantum-phase
 transition for incommensurate filling that reflects the competition between the two different coupling 
mechanisms using a phenomenological bosonization approach. 
The transition separates a superfluid phase from a CDW phase when the Luttinger parameter
of the bosons fields is $K=1/2$. This value can be reached here thanks to the long-range nature of the dipole-dipole interaction.
In stark contrast to a single-tube case where the transition is a simple cross-over, 
the main effect of coupling the 1D tubes through hopping and density interactions is to  give rise to a
genuine quantum-phase transition.

Using the powerful machinery of the CFT approach, we have determined the main 
long-wavelength properties of this transition which turns out to be very exotic.
In particular, we have revealed that the quantum phase transition
is described by the SU(2)$_N$ CFT, or equivalently 
to the U(1) $\times$ $\ZZ_N$ CFT, with a 
fractional central charge $c=3N/(N+2)$. The $\ZZ_N$ critical degrees of freedom 
are highly nontrivial and nonlocal with respect to the original atoms or polar molecules of the model.
The quantum critical point can be attained by a fine-tuning of the coupling constants that seem reasonably 
realistic in the context of cold polar molecules.
In this respect, this work opens the possibility to investigate the exotic physics of $\ZZ_N$
parafermions in the context of ultracold quantum bosonic gases. 

In the $N=2$ case, we have investigated in details the main characteristics 
of the quantum critical point which was shown to belong to the SU(2)$_2$ universality class
with central charge $c=3/2$. The triple-tube case ($N=3$)  is particularly 
promising since the quantum phase transition is governed by U(1)$\times\ZZ_3$ degrees of freedom.
In particular, within our low-energy approach, we have connected this problem to
the physics of the 2D $\ZZ_3$ chiral Potts model.
In this respect, it will be  important to carry out a thorough large-scale numerical analysis for mapping out the zero-temperature phase diagram of the triple-tube systems by adding an optical lattice in the $x$-direction. 
The commensurate case is also interesting and will be investigated elsewhere.
We hope that ongoing experimental research on polar molecules will allow to probe the quantum phase transition 
discussed in this paper.


\section*{Acknowledgements}
We would like to thank A. A. Nersesyan for encouragement and his interest in the work.
We are also grateful to E. Orignac and N. Zinner for hepful discussions.

\appendix

\section{Free-field representations}

In this Appendix, we present some important technical details on the SU(2)$_N$ quantum critical properties of the effective Hamiltonian (\ref{hamefftrans}), which controls the quantum phase transition of the coupled tubes system.

\subsection{Free-field representation of SU(2)$_N$ CFT}

As stated in Sec. II B, for $g_{\perp} = g_{d}$ and $K=1/2$, the Hamiltonian is given by Eq. (\ref{sdsg}) and displays an enlarged SU(2)$_L$ $\times$ SU(2)$_R$ global symmetry. Let us consider its non-interacting 
part ${\cal H}^{*}_0$, where ``$*$'' denotes the $g_{\perp} = g_{d}$ point. Each boson $\varphi_n$ 
describes an SU(2)$_1$ CFT with central charge $c=1$, generated by the left and right SU(2)$_1$ current:
\begin{equation}	
\begin{array}{lll}
	j^{\dagger}_{n R,L} &=&\ds \frac{1}{2 \pi} :\exp( \mp i \sqrt{8\pi}\; \varphi_{n R,L}):\\
	j^z_{n R,L} &=&\ds  \frac{1}{\sqrt{2 \pi}}   \partial_x \varphi_{n R,L} ,
\end{array}
\label{an:su21current}
\end{equation}	
where $:A:$ is the standard normal ordering of a bosonic operator $A$.
 In this work, the chiral bosonic fields $\varphi_{nR(L)}$ are defined as:
\begin{eqnarray}
	\varphi_n&=&\varphi_{nL}+\varphi_{nR},\nonumber\\
	\vartheta_n&=&\varphi_{nL}-\varphi_{nR},
\end{eqnarray}
and we are working with the prescription: $\left[\varphi_{nR}, \varphi_{mL} \right] = i \delta_{nm}/4$.
The bosonic fields are normalized by the following operator product expansion (OPE):
\begin{eqnarray}
	\varphi_{nL}(z)\varphi_{mL}(\omega)&\sim& - \frac{\delta_{nm}}{4\pi}\ln(z-\omega),\nonumber\\
	\varphi_{nR}(\bar{z})\varphi_{mR}(\bar{\omega})&\sim& - \frac{\delta_{nm}}{4\pi}\ln(\bar{z}-\bar{\omega}),
	\label{eq:OPEfreeboson}
\end{eqnarray}
with $z=v\tau + ix$, $\bar{z}=v\tau - ix$, and $\tau$ is the imaginary time. 

Eqs. (\ref{an:su21current}) give a free-boson representation of the SU(2)$_1$ currents that satisfy the so-called SU(2)$_1$ Kac-Moody algebra defined by the OPE:\cite{dms}
\begin{eqnarray}
j^{\alpha}_{n L} \left(z\right)
j^{\beta}_{n L} \left(\omega \right) \sim
\frac{ \delta^{\alpha \beta}}{8 \pi^2 (z-\omega)^2}
+\frac{i\epsilon^{\alpha \beta \gamma} 
j^{\gamma}_{ n L} \left(\omega \right)}{2 \pi (z-\omega)},
\label{app:curopesu2}
\end{eqnarray}
with $\alpha,\beta,\gamma = x,y,z$ and a similar result for the right current. 
Indeed, using Eq. (\ref{eq:OPEfreeboson}) and OPEs involving vertex operators:\cite{dms}
\begin{equation}
\begin{array}{l}
	\ds:e^{ia\varphi_{nL}}:(z):e^{ib\varphi_{nL}}:(\omega)\\
	\ds\qquad\sim (z-\omega)^{ab/4\pi} :e^{ia\varphi_{nL}(z)+ib\varphi_{nL}(\omega)}:,\\
	\partial_x \varphi_{n L}(z):e^{ia\varphi_{nL}}:(\omega)
	\ds\sim\frac{a}{4\pi (z-\omega)} :e^{ia\varphi_{nL}}:(\omega),
\end{array}
\label{app:OPEvertex}
\end{equation}
one shows that: 
\begin{equation}
\begin{array}{lll}
j^{\pm}_{n L}(z) j^{\mp}_{n L}(\omega)
	&\sim&\ds
	\frac{1}{4\pi^2 (z-\omega)^2}:e^{\pm i \sqrt{8\pi}(\varphi_{nL}(z)-\varphi_{nL}(\omega))}:,\\
	&\sim&\ds
	\frac{1}{4\pi^2 (z-\omega)^2}\pm\frac{1}{\pi (z-\omega)}j^{z}_{n L}(\omega),\\
j_{n L}^\pm(z) j^z_{n L}(\omega) 
	&\sim&\ds
	\frac{1}{2\pi\sqrt{2\pi}}
	\frac{\mp \sqrt{8\pi}}{4\pi (z-\omega)}:e^{\pm i\sqrt{8\pi}\varphi_{nL}}:(\omega),\\
	&\sim&\ds
	\frac{\mp 1}{2\pi (z-\omega)}j_{n L}^\pm(\omega),\\
j_{n L}^z(z) j_{n L}^z(\omega)
	&\sim&\ds
	\frac{1}{8\pi^2 (z-\omega)^2} .
\end{array}
\end{equation}
This is nothing else but the algebra (\ref{app:curopesu2}).

As any CFT, the SU(2)$_1$ CFT described by the bosonic fields $\varphi_{nR(L)}$ possesses a stress-energy tensor. Its left -- i.e. holomorphic -- component is given by:
\begin{equation}
	T_n=\frac{4 \pi^2}{3} : {\bf j}_{n L} \cdot {\bf j}_{n L} :,
\end{equation}
and satisfies the defining relation of a stress-energy tensor of a CFT: \cite{dms}
\begin{equation}
	T(z) T(\omega) \sim \frac{c/2}{(z-\omega)^4}
	+ \frac{2 T(\omega)}{ (z-\omega)^2} + \frac{\partial T(\omega)}{z-\omega},
\label{stress}
 \end{equation}
where $\pd=\partial/\partial z$, and $c$ is the central charge for the CFT in question; for the SU(2)$_1$ CFT, $c=1$ as already mentioned before. Using the representation (\ref{an:su21current}), the SU(2)$_1$ stress-energy tensor 
can be directly expressed in terms of the boson field:
\begin{equation}
	T_n= 2 \pi :(\partial_x\varphi_{nL})^2:;
\label{bosonstress}	
\end{equation}
it is straightforward to check that the definition (\ref{stress}) is indeed reproduced.

The non-interacting part of the Hamiltonian (\ref{sdsg}), ${\cal H}^{*}_0$,  can be written in terms of the
$N$ SU(2)$_1$ currents (\ref{an:su21current}):
\begin{equation}
{\cal H}^{*}_0 = \sum_{n=1}^{N} \int dx \; \frac{2 \pi v}{3} \left( : {\bf j}_{n L} \cdot {\bf j}_{n L} :
+ : {\bf j}_{n R} \cdot {\bf j}_{n R} : \right),
\end{equation}
and its underlying CFT is thus SU(2)$_1 \times $SU(2)$_1 \times \cdots \times$ SU(2)$_1$ with central charge $N$.
If we now combine the $N$ SU(2)$_1$ currents together:
\begin{eqnarray}	
I^{\dagger}_{R,L} &=& \frac{1}{2 \pi} \sum_{n=1}^{N} :\exp( \mp i \sqrt{8\pi} \varphi_{n R,L}):,
\nonumber\\
I^z_{R,L} &=&  \frac{1}{\sqrt{2 \pi}}  \sum_{n=1}^{N} \partial_x \varphi_{n R,L},
\label{app:su2Ncurrent}
\end{eqnarray}	
the resulting object is an SU(2)$_N$ current. Indeed, one can show, similarly to
the SU(2)$_1$ case, that it satisfies the SU(2)$_N$ Kac-Moody algebra:
\begin{eqnarray}
I^{\alpha}_{L} \left(z\right)
I^{\beta}_{L} \left(\omega \right) \sim
\frac{N \delta^{\alpha \beta}}{8 \pi^2 \left(z - \omega\right)^2}
+\frac{i\epsilon^{\alpha \beta \gamma}
I^{\gamma}_{ L} \left(\omega \right)}{2 \pi
\left(z - \omega\right)}.
\label{curope}
\end{eqnarray}
The holomorphic part of the stress-energy tensor of the SU(2)$_N$ CFT is defined by: 
\begin{equation}
	T_{\text{SU(2)}_N} = \frac{4 \pi^2}{N+2} :{\bf I}_L \cdot {\bf I}_L:.
\end{equation}
It satisfies the relation (\ref{stress}) with a central charge $c_{\text{SU(2)}_N}=3N/(N+2)$.
Using the identification (\ref{app:su2Ncurrent}), we find a free-field representation 
of the SU(2)$_N$ stress-energy tensor in terms of the bosonic fields $\varphi_{n L}$:
\begin{eqnarray}
	T_{\text{SU(2)}_N}&=&\frac{6\pi}{N+2}\sum_{n=1}^{N} :(\partial_x\varphi_{nL})^2:\nonumber\\
	&+&\frac{1}{N+2}\sum_{n\neq m=1}^N:\cos{[\sqrt{8\pi}(\varphi_{nL}-\varphi_{mL})]}:\nonumber\\
	&+&\frac{2\pi}{N+2}\sum_{n\ne m=1}^N :\partial_x\varphi_{nL}\partial_x\varphi_{mL}: .
	\label{app:SU2Nstress}
\end{eqnarray}

\subsection{Self-dual perturbation and ${\cal G}_N$ primary field}

In Sec. \ref{sec:confemb}, in order to investigate the critical properties of the model, we have 
introduced the ${\cal G}_N$ CFT through the conformal embedding (\ref{embedgenn}):
\begin{equation}
\text{SU(2)}_1 \times \text{SU(2)}_1 \times \cdots
\times \text{SU(2)}_1 \rightarrow \text{SU(2)}_N \times {\cal G}_N.
\label{app:embedgenn}
\end{equation}
Its stress-energy tensor is the difference between the $N$ SU$(2)_1$ and the SU$(2)_N$ stress-energy tensors, i.e., in the bosonic description:
\begin{eqnarray}
	T_{{\cal G}_N} 
	&=&\left(\sum_{n=1}^{N} T_{n}\right) - T_{SU(2)_N}\\
	&=&\frac{2\pi (N-1)}{N+2}\sum_n :(\partial_x\varphi_{nL})^2:\nonumber\\
	&-&\frac{1}{N+2}\sum_{n\neq m}:\cos{(\sqrt{8\pi}(\varphi_{nL}-\varphi_{mL}))}:\nonumber\\
	&-&\frac{2\pi}{N+2}\sum_{n \ne m} :\partial_x\varphi_{nL}\partial_x\varphi_{mL}: .
\label{app:GNstress}	
\end{eqnarray}
Let us show that indeed $T_{{\cal G}_N} $ satisfies the OPE of stress-energy tensors (\ref{stress})
with a central charge $c_{\text{G}_N}=N(N-1)/(N+2)$. Since we already know the OPE that $T_{n}$ and $T_{SU(2)_N}$ satisfy (see Eq. (\ref{stress})), we only need to compute the OPE between the $N$ SU$(2)_1$ and the SU$(2)_N$ stress-energy tensors. Using the free-field representations (\ref{app:SU2Nstress}, \ref{app:GNstress}) and 
Eq. (\ref{app:OPEvertex}), one obtains:
\begin{eqnarray}
T_{\text{SU(2)}_N}(z) \sum_{n=1}^{N}T_{n}(\omega)&\sim&	\frac{3N}{2(N+2)}\frac{1}{(z-\omega)^4}+\frac{2T_{\text{SU(2)}_N}(\omega)}{(z-\omega)^2}\nonumber\\
&+&\ds\frac{\partial T_{\text{SU(2)}_N}(\omega)}{z-\omega} .
\end{eqnarray}
The OPE $\sum_{n=1}^{N}T_{n}(z)T_{\text{SU(2)}_N}(\omega)$ can be easily deduced from the former by exchanging $z$ and $\omega$, then performing an expansion of $z$ around $\omega$; it has the exact same form. Combining all these OPEs together, we finally have:
\begin{eqnarray}
	&&T_{{\cal G}_N}(z)T_{{\cal G}_N}(\omega)\nonumber\\
	&=&
	\sum_{n,m=1}^{N} T_{n}(z)T_{m}(\omega) + T_{SU(2)_N}(z)T_{SU(2)_N}(\omega)\nonumber\\
	&&-T_{\text{SU(2)}_N}(z) \sum_{n=1}^{N}T_{n}(\omega)-\sum_{n=1}^{N}T_{n}(z)T_{\text{SU(2)}_N}(\omega)\nonumber\\
	&\sim&\frac{N(N-1)}{2(N+2)z^4}
	+ \frac{2 T_{{\cal G}_N}(\omega)}{ (z-\omega)^2} + \frac{\partial T_{{\cal G}_N}(\omega)}{z-\omega},
\end{eqnarray}
which proves the value of the central charge $c_{\text{G}_N}=N(N-1)/(N+2)$ asserted above.

In the analysis of Sec. II C, we have used some crucial properties of 
the self-dual term (\ref{sdgint}) that is responsible for the quantum 
phase transition in the dipolar bosons tubes. With the definitions above at hand, we now 
demonstrate these properties. In terms of the bosonic fields, the self-dual term reads:
\begin{eqnarray}
{\cal V}_{\rm sd} &=& \sum_{i=1}^{N-1} :\cos(\sqrt{2 \pi}(\varphi_{i+1} -  \varphi_{i})): \nonumber\\
&&\quad+  :\cos(\sqrt{2 \pi} (\vartheta_{i+1} -  \vartheta_{i})):.
\label{app:sdgint}
\end{eqnarray}
Let us check explicitely that it is a singlet under the SU(2)$_N$ CFT:
\begin{equation}
	{\vec I}_L(z)  {\cal V}_{\rm sd}(\omega,\bar{\omega})  \sim \vec 0,
\label{app:OPEsinglet}
\end{equation}
and similarly for the right component.
First of all, we have, using Eq. (\ref{app:OPEvertex}):
\begin{equation}
\begin{array}{l}
I^x_L (z):\cos\sqrt{2\pi}(\pkp-\pkm):(\omega,\bar{\omega})\\
	\sim\ds\frac{1}{4\pi}\sum_i\frac{-i}{(z-\omega)}
	\Big(\delta^{i,k+1}[\tC_{k+1}C_k-\tS_{k+1}S_k]\\
	\qquad\qquad+\delta^{i,k}[C_{k+1}\tC_k-S_{k+1}\tS_k]\Big)(\omega,\bar{\omega}),\\
		\sim\ds\frac{-i}{4\pi (z-\omega)}
		\Big(\tC_{k+1}C_k-\tS_{k+1}S_k \\
		+C_{k+1}\tC_k-S_{k+1}\tS_k\Big)(\omega,\bar{\omega}),\\
I^x_L(z):\cos(\sqrt{2 \pi} (\vartheta_{k+1} -  \vartheta_{k})):(\omega,\bar{\omega})\\
\sim\ds\frac{i}{4\pi (z-\omega)}
		\Big(C_{k+1}\tC_k-S_{k+1}\tS_k \\
		+\tC_{k+1}C_k-\tS_{k+1}S_k\Big)(\omega,\bar{\omega}),\\
\end{array}
\end{equation}
where we have adopted the shorthand notation:
\begin{equation}
	:\cos{(\sqrt{2\pi}\varphi_{k})}:=C_k, \qquad :\cos{(\sqrt{2\pi}\vartheta_{k})}:=\tC_k,
\end{equation}
and with similar definitions for the sine functions $S_k$ and $\tS_k$.  We conclude thus:
${I_L}^x(z)\cV_{\rm sd}(\omega,\bar{\omega})\sim 0$, and 
similarly for the $y$ component of the SU$(2)_N$ current $I_L^y$. For the $z$ component, we have, using Eq. (\ref{app:OPEvertex}):
\begin{equation}
\begin{array}{l}
	{I_L}^z(z):\cos\sqrt{2\pi}(\pkp-\pkm):(\omega,\bar{\omega})\\
	\sim\ds \frac{i}{4\pi}\sum_i\frac{1}{z-\omega}\Big(
	\delta^{i,k+1}:\sin\sqrt{2\pi}(\pkp-\pkm):\\
	\qquad\qquad\qquad -\delta^{i,k}:\sin\sqrt{2\pi}(\pkp-\pkm):
	\Big)(\omega,\bar{\omega})\\
	\sim 0.
\end{array}
\end{equation}
The same also applies for the second term of $\cV_{\rm sd}$ and that finally proves Eq. (\ref{app:OPEsinglet}).

The second important property is that ${\cal V}_{\rm sd}$ is a primary field under the ${\cal G}_N$ CFT, with scaling dimension $\Delta_{\cal V}=1$:
\begin{equation}
T_{{\cal G}_N} (z) {\cal V}_{\rm sd}(\omega,\bar{\omega}) \sim 
\frac{ {\cal V}_{\rm sd}(\omega,\bar{\omega})}{2 (z-\omega)^2} + \frac{\partial  {\cal V}_{\rm sd}(\omega,\bar{\omega})}{ z-\omega}.
\label{app:OPEVprimary}
\end{equation}
Let us check that $\cV_{\rm sd}$ displays such property. First,
we have just shown that $\vv{I_L}(z)\cV_{\rm sd}(\omega,\bar{\omega})\sim 0$, 
so it is straightforward to see that $T_{\text{SU(2)}_N}(z)\cV_{\rm sd}(\omega,\bar{\omega})\sim 0$. 
We are left with the OPE of the $N$ SU$(2)_1$ stress-energy tensors with ${\cal V}_{\rm sd}$:
\begin{equation}
\begin{array}{l}
	\ds\sum_{n=1}^{N} T_n(z):\cos[\sqrt{2 \pi}(\pkm-\pkp)]:(\omega,\bar{\omega})\\
	\ds\sim\sum_{n=1}^{N}
	\frac{1}{4}\frac{1}{(z-\omega)^2}\left(\delta_{n,k}+\delta_{n,k+1}\right)\\
	\ds\qquad\qquad\times:[C_{k} C_{k+1}+S_{k} S_{k+1}]:(\omega,\bar{\omega})\\
	\ds+\sum_{n=1}^{N}
	\frac{1}{z-\omega}\Bigg\{
	\delta_{n,k}\left[\pd(C_{k})C_{k+1}+\pd(S_{k})S_{k+1}\right]\\
	\ds\qquad\qquad\;+\delta_{n,k+1}\left[C_{k}\pd(C_{k+1})+S_{k}\pd(S_{k+1})\right]
		\Bigg\}(\omega,\bar{\omega}),\\
	\ds\sim
	\frac{1}{2}\frac{1}{(z-\omega)^2}
		:\cos\left[\sqrt{2 \pi}(\pkm-\pkp)\right]:(\omega,\bar{\omega})\\
	\ds\qquad+ \frac{1}{z-\omega}
		\pd\left(:\cos\left[\sqrt{2 \pi}(\pkm-\pkp)\right]:\right)(\omega,\bar{\omega}),
\end{array}
\end{equation}
and similarly for the second term of $\cV_{\rm sd}$, which proves finally Eq. (\ref{app:OPEVprimary}).

\subsection{Free-field representation of the TIM and $\ZZ_3$ Potts CFTs}

In this section, we  give a proof of the non-trivial identification (\ref{CFTidentities}) 
which occurs in triple-tube systems.

 It is first important to observe that the free massless bosonic
field $\Phi_2$ has a very special compactification radius
$R_2 =  \sqrt{3 /\pi}$ in the classification
of the CFT with central charge $c=1$.
At this radius, it displays a CFT with an extended
symmetry: a $N=2$ (respectively $N=1$) superconformal field theory (SCFT)
when the bosonic field $\Phi_2$ is compactified
along a circle (respectively an orbifold $\Phi_2 \sim - \Phi_2$) with radius $R_2$ \cite{waterson}.
Forgetting the contribution of the Majorana fermion $\xi^{2}_{L}$,
the conformal symmetry of the non-interacting limit of
model (\ref{bosoN3bis}) is Ising $\times$ [$c=1$ SCFT],
with central charge $c=3/2$. The latter CFT
can also be described in terms of the product
of TIM $\times$ $\ZZ_3$ Potts CFTs with central charge $c=7/10+4/5=3/2$. 
The precise conformal embedding has been derived
by the authors of Ref. \onlinecite{bul}:
\begin{eqnarray}
\ZZ_2 \times \left(c=1 \;  N=1 \;  {\rm SCFT} \right)
&=& P\left[{\cal M}_4 \times {\cal M}_5 \right],
\label{embeddingscft1}
\end{eqnarray}
where the projection $P$ restricts the primary operators of ${\cal M}_4 \times {\cal M}_5$
to the subset: $\{\Phi^{(4)}_{r,s} \Phi^{(5)}_{s,q} \}$,
$\Phi^{(p)}_{r,s}$ ($ 1 \le r \le p-1, 1 \le s \le p$) being the primary operator of ${\cal M}_p$ CFT.

As is well-known, the (left) stress-energy tensor of the $\ZZ_2$ (or Ising) CFT expresses directly in terms of the Majorana
fermion $\xi^{1}_{L}$:
$T_I = -\pi : \xi_L^{1} \partial \xi_L^{1}:$. It is indeed straightforward to check that $T_I$ satisfies 
the defining relation (\ref{stress}) of a stress-energy tensor with $c=1/2$. The left Majorana
$\xi^{1}_{L}$ is a primary field with holomorphic weight $h=1/2$:
\begin{eqnarray}
T_I \left(z\right) \xi^{1}_{L} \left(0\right) &\sim& \frac{ \xi^{1}_{L} \left(0\right)}{2 z^2 }
+  \frac{ \partial \xi^{1}_{L} \left(0\right)}{z } \nonumber \\
 \xi^{1}_{L} \left(z\right) \xi^{1}_{L} \left(0\right) &\sim& \frac{1}{2 \pi z} + \frac{z}{\pi} T_I \left(0\right)
 + \frac{z^2}{2\pi} \partial T_I \left(0\right).
\label{majodefope}
\end{eqnarray}
For the $N=1$ SCFT with central charge $c=1$, the stress-energy tensor is simply given
by Eq. (\ref{bosonstress}):  $T_0 = - 2  \pi :(\partial\Phi_{2L})^2: $.
On top of this bosonic tensor, the $N=1$ SCFT is characterized by the existence
of a fermionic current $G$ with holomorphic weight $h=3/2$: \cite{friedan}
\begin{eqnarray}
T_0 \left(z\right) G \left(0\right) &\sim& \frac{ 3 G \left(0\right)}{2 z^2 }
+  \frac{ \partial G \left(0\right)}{z } \nonumber \\
G \left(z\right) G \left(0\right) &\sim& \frac{1}{z^3} +  \frac{ 3  T_0 \left(0\right)}{z }
+  \frac{ 3 \partial  T_0 \left(0\right)}{2 } .
\label{scftope}
\end{eqnarray}
In the particular case of the $N=1$ SCFT with central charge $c=1$, the current $G$ has a simple free-field
representation in terms of the free-massless boson $\Phi_{2L}$:\cite{waterson}
\begin{eqnarray}
G = \sqrt{2} :\cos(\sqrt{12 \pi} \Phi_{2L}):.
\label{susycurrent}
\end{eqnarray}
It is indeed easy to check, using Eq. (\ref{app:OPEvertex}), that the OPEs (\ref{scftope}) are 
reproduced from this identification.

From all these definitions, we find a new free-field representation of the 
left stress-energy tensors of the TIM and $\ZZ_3$ Potts CFTs in terms of  a Majorana
fermion and a free-massless boson:
\begin{eqnarray}
	 T_{\rm TIM}&=&\frac{1}{5} T_I + \frac{3}{5} T_0 -  \frac{4\sqrt{\pi}}{5}   \xi^{1}_{L} :\cos(\sqrt{12 \pi} \Phi_{2L}):\quad\quad\nonumber\\
	 T_{\ZZ_3} &=&\frac{4}{5} T_I + \frac{2}{5} T_0 +  \frac{4\sqrt{\pi}}{5}   \xi^{1}_{L} :\cos(\sqrt{12 \pi} \Phi_{2L}):,
	\label{stressTIMPotts}
\end{eqnarray}
where $T_{\rm TIM}$ and $T_{\ZZ_3}$ denote respectively the stress-energy tensor of the TIM and
 $\ZZ_3$ Potts CFTs.
Using the results (\ref{majodefope}, \ref{scftope}), one can indeed show that $T_{\rm TIM}$ (respectively
$T_{\ZZ_3}$) satisfies the definition (\ref{stress}) with $c=7/10$ (respectively $c=4/5$),
together with the decoupling: $ T_{\rm TIM} (z) T_{\ZZ_3} (0) \sim 0$.
In addition, we note that we have $T_{\rm TIM} + T_{\ZZ_3} =  T_I + T_0$ in full agreement with
the existence of the conformal embedding (\ref{embeddingscft1}).

We are now in position with the free-field representation (\ref{stressTIMPotts}) to show the
identification (\ref{CFTidentities}). In this respect, let us introduce the following operators:
\begin{eqnarray}
{\cal O}_1 &=& \mu_1 :\cos\left(\sqrt{3 \pi} \Phi_2\right):  +
\sigma_1 :\cos\left(\sqrt{3 \pi} \Theta_2\right):\quad\quad 
\nonumber \\
{\cal O}_2 &=& \mu_1 :\cos\left(\sqrt{3 \pi} \Phi_2\right):  -
\sigma_1 :\cos\left(\sqrt{3 \pi} \Theta_2\right): .
\label{defop}
\end{eqnarray}
Let us first show that ${\cal O}_1$ is a singlet under the $\ZZ_3$ Potts CFT and a primary
operator of the TIM CFT with holomorphic weight $h=7/16$.
To this end, we need the following OPEs for a massless boson field that can be obtained
from Eq. (\ref{app:OPEvertex}):
\begin{equation}
\begin{array}{ll}
\ds:\cos(\sqrt{12 \pi} \Phi_{2L}): \left(z\right) :\cos(\sqrt{3 \pi} \Phi_{2}): \left(0,0\right) \sim& \\
\ds\frac{e^{-i 3 \pi/4}}{2 z^{3/2}}  \left[ :\cos(\sqrt{3 \pi} \Theta_{2}): 
	 + 2 z   :\partial \cos(\sqrt{3 \pi} \Theta_{2}): \right] \left(0,0\right) \\
\ds:\cos(\sqrt{12 \pi} \Phi_{2L}): \left(z\right) :\cos(\sqrt{3 \pi} \Theta_{2}): \left(0,0\right) \sim& \\
\ds\frac{e^{i 3 \pi/4}}{2 z^{3/2}} \left[ :\cos(\sqrt{3 \pi} \Phi_{2}): 
	 + 2 z  :\partial \cos(\sqrt{3 \pi} \Phi_{2}): \right] \left(0,0\right) .
\end{array}
	\label{opegaussiansusy} 
\end{equation}
The following Ising OPEs, which can be found in Ref. \onlinecite{zamolochatterjee}, are also needed:
\begin{eqnarray}
 \xi^{1}_{L} \left(z\right) \sigma_1\left(0,0\right) \sim 
	\frac{e^{i \pi/4}}{2 \sqrt{\pi} z^{1/2}}  \left[  \mu_1
	 + 4 z  \partial   \mu_1 \right] \left(0,0\right)\nonumber \\
 \xi^{1}_{L} \left(z\right) \mu_1\left(0,0\right) \sim 
	\frac{e^{-i \pi/4}}{2 \sqrt{\pi} z^{1/2}}  \left[  \sigma_1
	 + 4 z  \partial   \sigma_1 \right] \left(0,0\right) .
\label{opeisingzamolo} 
\end{eqnarray}
With all these identifications, it becomes straightforward to obtain the following result directly 
from Eq. (\ref{stressTIMPotts}):
\begin{eqnarray}
T_{\rm TIM} (z) {\cal O}_1(0,0) &\sim& 
\frac{ 7 {\cal O}_1(0,0)}{16 z^2} + \frac{\partial {\cal O}_1 (0,0)}{ z} 
\nonumber \\
T_{\ZZ_3} (z) {\cal O}_1(0,0) &\sim& 0 ,
\label{TIMprimary}
\end{eqnarray}
which states that ${\cal O}_1$ is indeed a primary operator of the TIM CFT with
scaling dimension $7/8$. It is thus proportional to the subleading magnetization $\sigma^{'}_{\rm TIM}$.
Using the correlation (\ref{isingcorrelation}), one can fix the normalization factor by evaluating
the two-point function of ${\cal O}_1$:
\begin{eqnarray}
\langle {\cal O}_1(z,\bar z)  {\cal O}_1(0,0)  \rangle &\sim&   |z|^{-7/4}\nonumber\\
&=& \langle \sigma^{'}_{\rm TIM} (z,\bar z)\;  \sigma^{'}_{\rm TIM} (0,0)\rangle,\qquad
\label{sigmaTIM}
\end{eqnarray}
so that we deduce ${\cal O}_1 \sim \sigma^{'}_{\rm TIM} $.
Finally, from its definition (\ref{defop}),  
it is easy to check that ${\cal O}_2$ is a primary operator of the ${\cal M}_4 \times {\cal M}_5$ CFT
with scaling dimension $7/8$:
\begin{eqnarray}
\left(T_{\rm TIM} + T_{\ZZ_3} \right) (z) {\cal O}_2(0,0)
	&=&\ds\left(T_{I} + T_{0} \right) (z) {\cal O}_2(0,0)\\
	&\sim&\ds\frac{ 7 {\cal O}_2(0,0)}{16 z^2} + \frac{\partial {\cal O}_2 (0,0)}{ z}.\quad\quad\nonumber
\label{Tprimary}
\end{eqnarray}
The $P[{\cal M}_4 \times {\cal M}_5]$ CFT contains only two primary
operators with scaling dimension $7/8$, i.e.,  $\sigma^{'}_{\rm TIM}$ and $\sigma_{\rm TIM} \epsilon_{\ZZ_3}$.
We thus find that $ {\cal O}_2 \sim \sigma_{\rm TIM} \epsilon_{\ZZ_3}$ since ${\cal O}_1 \sim \sigma^{'}_{\rm TIM} $.

\end{document}